\documentclass[conference]{IEEEtran}
\IEEEoverridecommandlockouts
\usepackage{cite}
\usepackage{amsmath,amssymb,amsfonts}
\usepackage{algorithmic}
\usepackage{graphicx}
\usepackage{textcomp}
\usepackage[usenames,dvipsnames,table]{xcolor}
\usepackage{multirow}
\usepackage{tabularx}

\usepackage{hyperref}
\usepackage{booktabs}
\usepackage{makecell}
\usepackage[frozencache=true,cachedir=minted-cache]{minted} 
\usepackage{svg}
\newcommand{\ie}{{\em i.e.}, }
\newcommand{\eg}{{\em e.g.}, }
\newcommand{\shs}[1]{\shortstack[c]{#1}}

\usepackage[ruled, linesnumbered, noend]{algorithm2e}

\newcommand\blfootnote[1]{%
  \begingroup
  \renewcommand\thefootnote{}\footnote{#1}%
  \addtocounter{footnote}{-1}%
  \endgroup
}

\usepackage{subcaption}

\newcommand\newsubcap[1]{\phantomcaption%
       \caption*{\figurename~\thefigure\thesubfigure: #1}}

\newcommand{\pfvs}{\vspace{-0.6mm}}
\newcommand{\vspacesections}{\vspace{-0.0mm}}
\newcommand{\mysection}[1]{\vspacesections \section{#1} \vspacesections}
\newcommand{\mysubsection}[1]{\vspacesections \subsection{#1} \vspacesections}

\newcommand{\svgcap}{}

\def\BibTeX{{\rm B\kern-.05em{\sc i\kern-.025em b}\kern-.08em
    T\kern-.1667em\lower.7ex\hbox{E}\kern-.125emX}}
\begin{document}

\title{{CiMLoop}: A Flexible, Accurate, and Fast Compute-In-Memory Modeling Tool}

\author{
\IEEEauthorblockN{Tanner Andrulis}
\IEEEauthorblockA{\textit{MIT} \\ Cambridge, USA \\ andrulis@mit.edu}
\and
\IEEEauthorblockN{Joel S. Emer}
\IEEEauthorblockA{\textit{MIT, Nvidia} \\ Cambridge, USA \\ jsemer@mit.edu}
\and
\IEEEauthorblockN{Vivienne Sze}
\IEEEauthorblockA{\textit{MIT} \\ Cambridge, USA \\ sze@mit.edu}
}

\maketitle

\begin{abstract}
Compute-In-Memory (CiM) is a promising solution to accelerate Deep Neural Networks (DNNs) as it can avoid energy-intensive DNN weight movement and use memory arrays to perform low-energy, high-density computations. These benefits have inspired research across the CiM stack, but CiM research often focuses on only one level of the stack (\textit{i.e.,} devices, circuits, architecture, workload, or mapping) or only one design point (\textit{e.g.,} one fabricated chip). There is a need for a full-stack modeling tool to evaluate design decisions in the context of full systems (\textit{e.g.,} see how a circuit impacts system energy) and to perform rapid early-stage exploration of the CiM co-design space.

To address this need, we propose CiMLoop: an open-source tool to model diverse CiM systems and explore decisions across the CiM stack. CiMLoop introduces \textbf{(1)} a flexible specification that lets users describe, model, and map workloads to both circuits and architecture, \textbf{(2)} an accurate energy model that captures the interaction between DNN operand values, hardware data representations, and analog/digital values propagated by circuits, and \textbf{(3)} a fast statistical model that can explore the design space orders-of-magnitude more quickly than other high-accuracy models.

Using CiMLoop, researchers can evaluate design choices at different levels of the CiM stack, co-design across all levels, fairly compare different implementations, and rapidly explore the design space.
\end{abstract}


\mysection{Introduction}

Compute-In-Memory (CiM) is a promising solution to address the high data movement energy and large number of computations required by Deep Neural Networks (DNNs). CiM systems compute directly inside memory, letting them \textbf{(1)} keep DNN weights in memory to reduce high-energy data movement, and \textbf{(2)} use large memory arrays to compute many parallel multiply-accumulate (MAC) operations with high density and low energy.
\blfootnote{\label{cimloop_url} The CiMLoop source, documentation, and tutorials are available at \href{https://github.com/mit-emze/cimloop}{\(\textit{https://github.com/mit-emze/cimloop}\)}.}

Many recent CiM implementations explore different levels of the CiM stack: \textit{Devices} store weights in CiM arrays; \textit{Circuits} perform computations, analog/digital conversion, data movement, and other actions; \textit{Architecture} organizes devices, circuits, and other components into a larger system; \textit{Workloads} are the DNNs to accelerate; and \textit{Mapping} schedules workloads spatially and temporally on hardware.

Often, researchers explore only one of these levels~\cite{NVMExplorer,isaac}. However, CiM stack levels interact, so it is important to have a \textit{full-stack model}, which looks at all levels together. Full-stack modeling is essential because \textit{evaluating a choice at one level requires the context of the other levels}. For example, when choosing circuits, we would like to know how they will affect full-system energy/throughput when running a given workload. For this reason, \textit{we must co-optimize across the full stack to find the design that best meets desired criteria}.


To enable full-stack modeling, we need a modeling tool that can represent all levels of the CiM stack and how they interact with each other. To do so, such a tool must be \textit{flexible} to model the space of design decisions, \textit{accurate} to correctly compare these decisions, and \textit{fast} to explore the design space quickly. Table~\ref{tab:prior_works} compares CiMLoop to prior modeling works in addressing the following key challenges.

\def \g[#1]{\cellcolor{green!20}{#1}}
\def \m[#1]{\cellcolor{orange!20}{#1}}
\def \b[#1]{\cellcolor{red!20}{#1}}

\begin{table}
\centering \footnotesize
\begin{tabular}{ccccc}
\shs{Modeling\\Work} & \shs{Architecture\\Flexibility} & \shs{Circuit\\Flexibility} & \shs{Energy\\Accuracy} & \shs{Model\\Speed} \\
\midrule
NeuroSim~\cite{neurosim_1,neurosim_2,neurosim_3,neurosim_most_recent} & \b[Low] & \b[Low] & \g[High] & \b[Low] \\
MNSim~\cite{mnsim_1, mnsim_most_recent} & \b[Low] & \b[Low] & \b[Low] & \m[Med.] \\
Timeloop~\cite{accelergy,accelergy_pim,sparseloop_1,sparseloop_2,timeloop,ruby} & \g[High] & \b[Low] & \b[Low] & \g[High] \\
\textbf{This Work} & \g[\textbf{High}] & \g[\textbf{High}] & \g[\textbf{High}] & \g[\textbf{High}] \\
\end{tabular}
\caption{Comparison to prior CiM modeling works.} \pfvs \pfvs
\label{tab:prior_works}
\end{table}

\subsection{CiM Modeling Challenges}
\subsubsection{Flexibility Challenge} The modeling tool must model the CiM design space, but different CiM implementations introduce different circuits and architectures~\cite{NeuroSim_Validated,jia,sinangil,wan,wan_ii,wang,wang_ii,colonnade,dong_sram_cim,xue202015}. To model the design space, the tool must let users easily define circuits, architectures, and how data moves between them.


This is a challenge because data movement can be different between architectures (\eg SRAM buffers in the memory hierarchy store and exchange data with each other) and circuits (\eg within an SRAM buffer, sense amplifiers read and propagate signals from the SRAM array). To address this challenge, the modeling tool must be \textit{flexible}, meaning that it lets users easily describe a wide range of components, how they connect, and how they move data. Unfortunately, prior CiM modeling tools are either inflexible~\cite{neurosim_most_recent,mnsim_most_recent}, or lack circuit-level modeling~\cite{accelergy,timeloop}.

Solving this issue introduces a second challenge: Flexible modeling tools must describe many different types of components, including both an architecture hierarchy (\eg DRAM + L3/L2/L1 cache) and set of circuits (\eg data converters, SRAM bitcells, addressing circuitry). To address this, the modeling tool must also let users easily describe complex designs with many different types of components. 

\subsubsection{Accuracy Challenge} The tool must model energy accurately to correctly inform design decisions, but device and circuit energy is \textit{data-value-dependent}, meaning affected by the values of data that each component propagates (\eg energy of some ReRAM devices are proportional to computed MAC values~\cite{neurosim_most_recent})  Data-value-dependence can significantly affect overall system energy~\cite{sinangil}, so we need a modeling tool with high \textit{energy accuracy}, which we define as capturing data-value-dependence. Unfortunately, some prior models may use inaccurate fixed-energy or fixed-power models~\cite{accelergy, timeloop, mnsim_most_recent} that do not model data-value-dependence.

\subsubsection{Modeling Speed Challenge} The modeling tool must be fast enough to explore the large design space, but accurate data-value-dependent energy modeling depends on data values, and it is slow to simulate the many data values propagated by components in a CiM system. Unfortunately, prior CiM modeling works have not addressed this problem; accurate prior models are slow~\cite{neurosim_most_recent} and fast prior models are inaccurate~\cite{accelergy,accelergy_pim,sparseloop_1,sparseloop_2,timeloop,ruby, mnsim_most_recent}.



\subsection{CiMLoop}
To address these challenges, we propose CiMLoop: a full-stack CiM modeling tool with flexible user-defined systems and fast, accurate statistical energy modeling. CiMLoop makes the following key contributions:

\begin{itemize}
\item A flexible specification to describe CiM systems with user-defined circuits and architecture. This specification includes \textbf{(1)} directives representing circuit and architecture data movement patterns, and \textbf{(2)} a representation that describes both circuits and architecture in a single hierarchy. This hierarchy lets users easily define and map workloads to systems with many different types of circuits, architectures, and data movement patterns.

\item An accurate model of data-value-dependent component energy that captures the interactions between DNN operand values, data representations, and data values propagated by components. Using this interface, we develop a suite of CiM component models that can easily be used in user-defined systems.

\item Fast statistical models to calculate data-value-dependent energy. These models calculate average energy once for each action by each type of component, letting CiMLoop \textbf{(1)} use constant runtime to model an arbitrary number of components/actions, and \textbf{(2)} amortize energy calculation over many mappings. As a result, CiMLoop is orders-of-magnitude faster than prior accurate modeling works.

\item Case studies using four recently-published fabricated CiM designs. In these case studies, we validate CiMLoop's accuracy and show that CiMLoop can \textbf{(1)} model diverse CiM implementations, \textbf{(2)} explore tradeoffs at different levels of the CiM stack, \textbf{(3)} model full systems, and \textbf{(4)} fairly compare different CiM designs.

\end{itemize}

By addressing these challenges, CiMLoop lets researchers evaluate and co-design across levels, fairly compare CiM implementations, and rapidly explore the CiM design space.

\mysection{Background/Motivation} \label{bg_motivation}
We first give background on \textit{Compute-In-Memory (CiM)}  for accelerating DNNs and why full-stack modeling is essential for evaluating and exploring the CiM space. Following this, we discuss each of the three challenges addressed by CiMLoop.

\mysubsection{Compute-In-Memory (CiM) for DNNs}~\label{CiM_for_DNNs}
Tensor operations in Deep Neural Network (DNN) layers can use large input, weight, and output tensors that are energy-intensive to move from memory. Furthermore, tensor operations may require large numbers of multiply-accumulate (MAC) operations, which can also be energy and latency-intensive to compute~\cite{dnn_scaling,efficient_processing_of_dnns, horowitz}. 

Data movement consumes a significant portion of DNN energy~\cite{horowitz, eyeriss}, and CiM can reduce energy costs by computing directly within the memory arrays~\cite{PRIME}, which we define as two-dimensional grids of interconnected memory cells (\eg SRAM bitcells or RRAM devices). Most commonly, CiM systems keep DNN weights in memory because they do not change during DNN inference~\cite{PRIME,isaac,wang,wan}. Many CiM implementations further reduce energy and latency with analog domain computation, where an analog MAC operation can be completed with one or a few memory cells rather than a full digital MAC circuit.

\begin{figure}
\centering
\includegraphics[width=\linewidth, keepaspectratio] {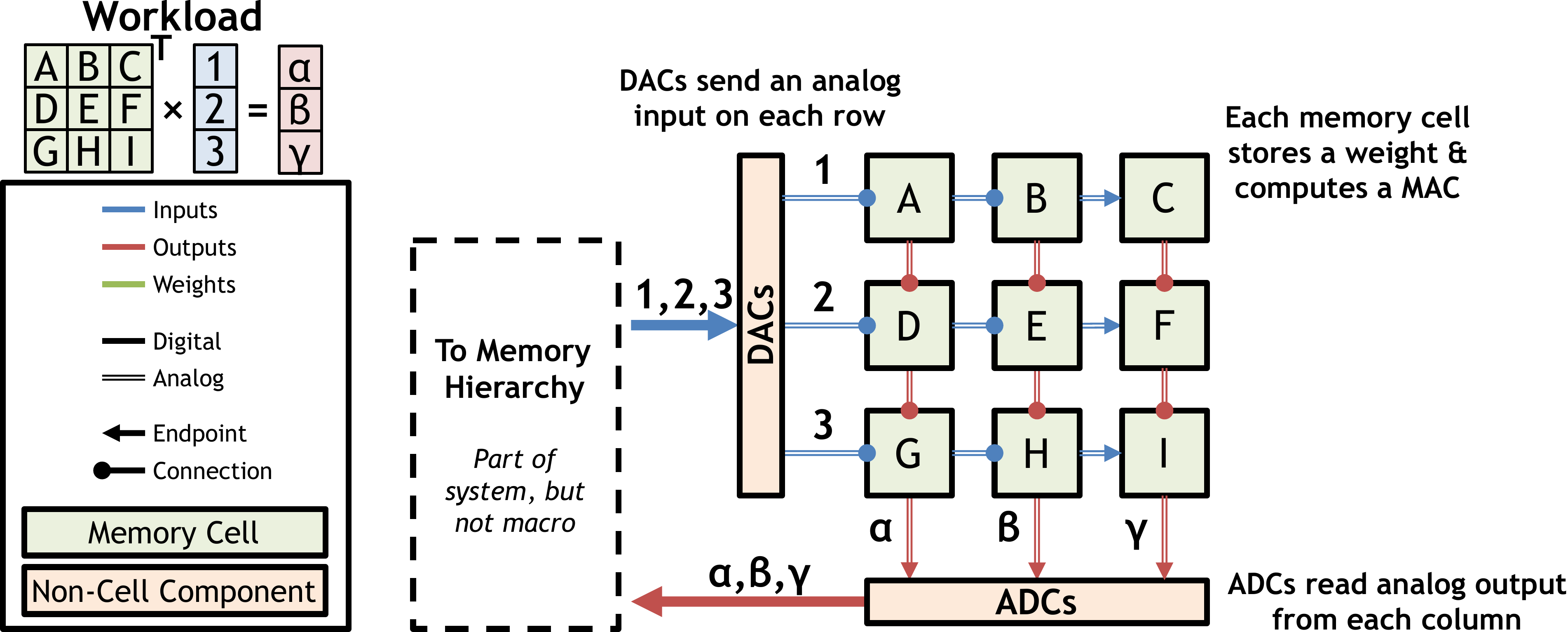}
\caption{CiM macro computing a matrix-vector multiplication. From a workload (top-left), the weight matrix (green) is programmed into memory cells. Input vector elements (blue) are sent on rows and outputs (red) appear on columns.}
\label{fig:how_cim_works}
\pfvs \end{figure}

Often, a CiM implementation is published as a \textit{macro}~\cite{jia,sinangil,wan,wan_ii,wang,wang_ii,colonnade,dong_sram_cim,xue202015}, which we define as an array of memory cells plus the additional components needed to compute full MAC operations. This is to contrast from the \textit{system}, which we define as one or more macros plus a memory hierarchy and the interconnects between memories in the hierarchy.

One commonly-explored macro topology~\cite{isaac,PRIME,dot_product_engine} is shown in Fig.~\ref{fig:how_cim_works}. In this topology, a digital-analog converter (DAC) supplies analog inputs to the rows of a memory array. Each \textit{memory cell} in the array computes an analog MAC operation between the input and its stored weight. Analog outputs from each memory cell in a column are summed and read by an analog-digital-converter (ADC) to yield a digital output. 

Researchers have explored a wide variety of other CiM macro topologies, circuits, and means of analog and/or digital computation. CiMLoop models the CiM macro space and integrates CiM macros with system modeling tools~\cite{accelergy,accelergy_pim,sparseloop_1,sparseloop_2,timeloop,ruby}.

\mysubsection{Full-Stack Modeling}
CiM research spans each of the following levels:
\begin{itemize}
    \item \textit{Devices}: The components forming each memory cell. Published macros often use SRAM~\cite{sinangil,wang}, DRAM~\cite{compute_in_dram, compute_in_dram_ii}, ReRAM~\cite{wan, dot_product_engine, ibm_14nm_pcm}, or STT-RAM~\cite{sttram_cim}.
    \item \textit{Circuits}: The components performing computation, analog/digital conversion, storage, data movement, and other actions~\cite{wang,sinangil,chen2022bit}.
    \item \textit{Architecture}: The organization of components into a larger system (\eg the number of each component and how components are connected)~\cite{CASCADE,PRIME,AtomLayer,raella,1T2R_Aeris,TIMELY,isaac}.
    \item \textit{Workload}: The DNN to be run, which we model as a series of extended-Einsum~\cite{teaal} operations with tensors of varying shapes and values~\cite{efficient_processing_of_dnns,timeloop}.
    \item \textit{Mapping}: The temporal and spatial scheduling of the workload onto the system~\cite{timeloop}.
\end{itemize}

\begin{figure}[]
    \centering
    \begin{subfigure}{\linewidth}
    \svgcap \includegraphics[width=\linewidth, keepaspectratio] {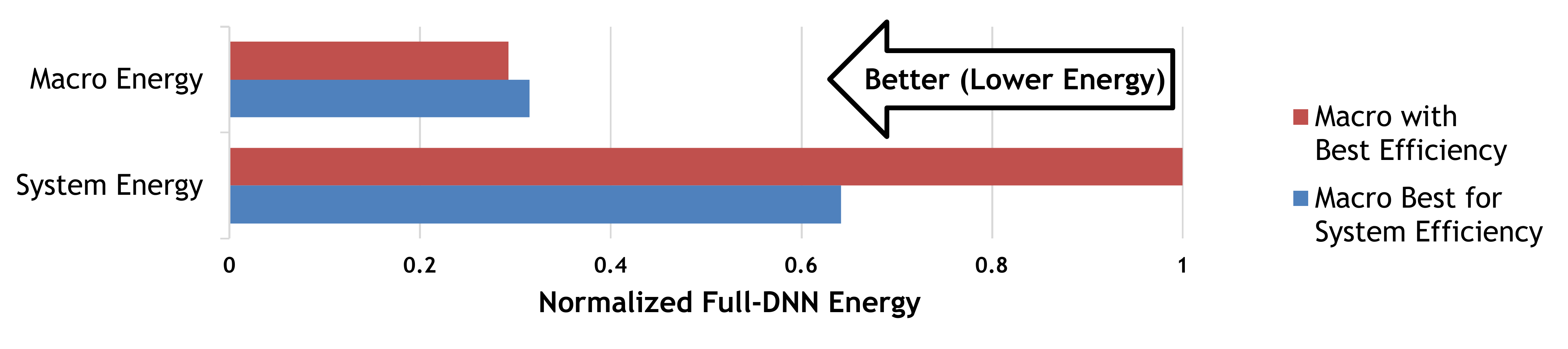}
    \newsubcap{\label{fig:why_full_stack_a} Optimizing for the lowest-energy macro while neglecting the system yields a higher-energy system overall.}
    \end{subfigure}
    \begin{subfigure}{\linewidth}
    \svgcap \includegraphics[width=\linewidth, keepaspectratio] {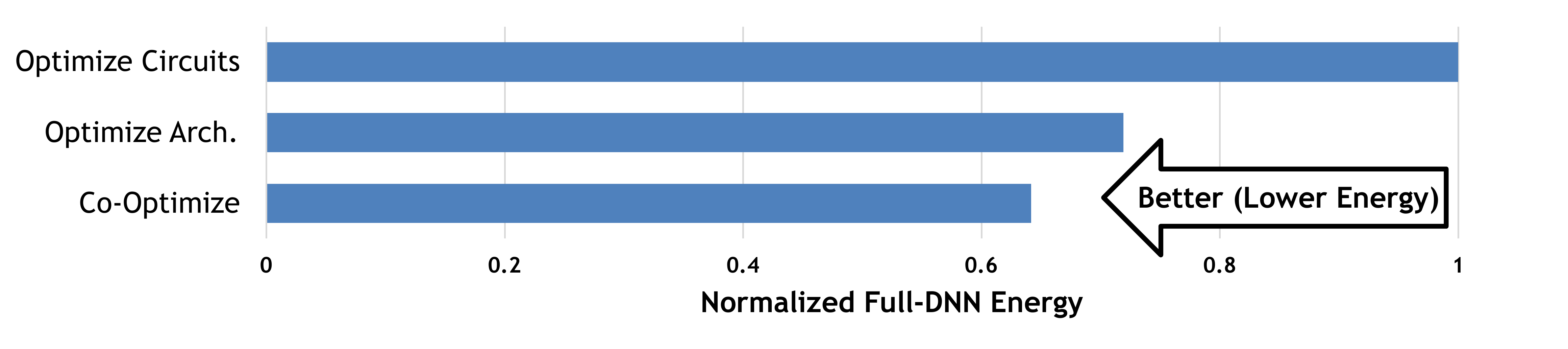}
    \newsubcap{\label{fig:why_full_stack_b} Co-optimizing circuits and architecture yields a lower-energy system than optimizing either individually.}
    \end{subfigure}
    
\pfvs \end{figure}

\textit{Full-stack modeling}, meaning modeling all levels together, is essential for two reasons. First, \textit{all-level context is needed to evaluate choices at a given level.} The pitfall of modeling without considering the full stack is shown in Fig.~\ref{fig:why_full_stack_a} where we explore different CiM array sizes for a macro running the DNN ResNet18~\cite{ResNet}. The \textbf{lowest-energy macro} has a smaller array, which maintains high utilization and low energy even with small DNN tensors. The \textbf{macro that yields the lowest-energy system} uses a larger array. Though the large array is often underutilized, it stores store more weights, letting it reduce data movement to/from the memory hierarchy to reduce system energy.

If we had optimized for macro energy, we would have been misled into a higher-energy system; only by considering the full system can we make the best decisions for a level.~\footnote{Furthermore, to conduct this exploration, we had to consider the workload (tensor sizes for each DNN layer) and mapping (maximizing array utilization).}


The second reason for full-stack modeling is that \textit{co-exploring levels can find better systems}~\cite{mars,forms,PIM-Prune,learning_sparsity_for_reram}. In Fig.~\ref{fig:why_full_stack_b}, we start with the lowest-energy macro from Fig.~\ref{fig:why_full_stack_a} and measure full-system energy while varying the DAC resolution (circuits) and CiM array size (architecture). The \textbf{optimize circuits} macro used a high-resolution DAC to process more input bits at a time, reducing the number of times the array had to be activated and decreasing energy per MAC. The \textbf{optimize architecture} macro, in addition to the high-resolution DAC, used a larger array to further reduce the number of times the array had to be activated for a given number of MACs. For smaller tensors, the array was underutilized, and high-resolution high-energy DACs resulted in a greater energy efficiency loss. Finally, the \textbf{optimize both} macro used a larger array and a low-resolution DAC. The large array could compute many MACs per activation, while the low-resolution DAC maintained low energy when the large array was underutilized. Here, co-optimization helped us find design decisions that synergized to produce a lower-energy system.

\mysubsection{CiM Circuits, Architectures, and Data Movement} 





A CiM macro consists of a set of components (\ie devices, circuits, and architecture) and the connections they use to move data. Component and data movement choices determine system area/energy and the computations that a system may carry out, so it is critical to represent these choices when modeling CiM macros.

Different CiM macros use different components and data movement patterns to optimize for different design goals. A key aspect to consider when choosing components and data movement patterns is how they can leverage \textit{data reuse} to reduce energy~\cite{eyeriss}. Reuse saves energy by reducing the number of times components are used to move and/or process data. For example, weights may be stored and reused temporally (\ie across cycles) in memory cells to reduce the number of times weights are fetched from separate memories. As another example, the macro in Fig.~\ref{fig:how_cim_works} spatially reuses analog inputs between multiple array columns, letting it reuse each analog input for multiple computations. This reduces the number of times that DACs convert inputs to analog.



There are many interesting design choices involving different components, data movement patterns, and reuse opportunities. To illustrate some of such choices, Fig.~\ref{fig:example_macros} shows several strategies that published macros use to reduce ADC energy. In the \textit{Base Macro}~\cite{NeuroSim_Validated}, multiple array rows reuse outputs (\ie an $N$-row array may sum outputs from up to $N$ MACs and read the result with one ADC convert)\footnote{Many works have introduced strategies to reduce ADC energy. For more information, see the Titanium Law~\cite{raella}, which breaks down the factors that contribute to ADC energy and shows how ADC energy can be reduced.}. To further reduce ADC energy, different macros use a variety of strategies:
\newcolumntype{Y}{>{\centering\arraybackslash}X}
\newcolumntype{Z}{>{\arraybackslash}X}
\begin{figure*}
\centering
    \centering
    \svgcap \includegraphics[width=\linewidth, keepaspectratio] {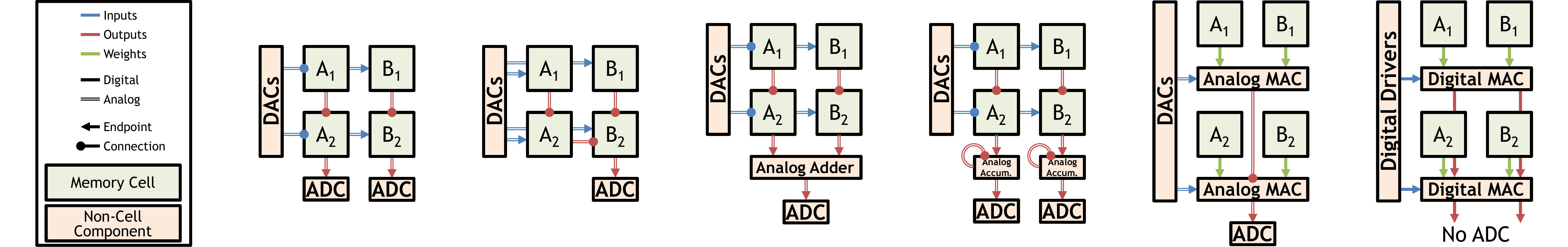}
    
\vspace{-30mm}
{\footnotesize 
\addtolength{\tabcolsep}{-0.8em}
\begin{tabularx}{\textwidth}{Y|Y|Y|Y|Y|Y|Y}
&&&&&&\\
&&&&&&\\
&&&&&&\\
&&&&&&\\
&&&&&&\\
&&&&&&\\
&&&&&&\\
&&&&&&\\
&&&&&&\\
&&&&&&\\
 & \textbf{\textit{Base Macro~\cite{NeuroSim_Validated}}} & \textbf{\textit{A}~\cite{jia}} & \textbf{\textit{B}~\cite{sinangil}} & \textbf{\textit{C}~\cite{wan,wan_ii}} & \textbf{\textit{D}~\cite{wang,wang_ii}} & \textbf{\textit{Digital CiM}~\cite{colonnade}} \\
\hline
\textit{From where} outputs are reused (summed)  & Different rows & Different rows and \textbf{columns} & Different rows and \textbf{columns} & Different rows and \textbf{cycles} & Different rows and \textbf{columns} & Different rows \\
\hline
\textit{How} outputs are reused (summed) & Wire & Wire & Wire and \textbf{analog~adder} & Wire and \textbf{analog~accumulator} & Wire and \textbf{analog~MAC} & \textbf{Digital~MAC} \\
\hline
Analog or digital & Analog & Analog & Analog & Analog & Analog & \textbf{Digital} \\
\hline
Benefit of strategy & - & \multicolumn{4}{c|}{$\uparrow$ Output reuse, $\downarrow$ ADC converts} & Eliminate ADC \\
\hline
Cost of strategy & - & $\downarrow$~Input reuse, $\uparrow$~DAC converts & \multicolumn{4}{c}{Area and energy of extra components.} \\
\hline
Mapping restriction of~strategy & - & $A_i/B_i$ store bits of different weights & $A_i/B_i$ store different bits of same weight & Different input bits in consecutive cycles & $A_i/B_i$ store different bits of same weight & $A_i/B_i$ store different bits of same weight
\end{tabularx}
}
    \caption{ADC-energy-reducing strategies of published CiM macros~\cite{NeuroSim_Validated,jia,sinangil,wan,wan_ii,wang,wang_ii,colonnade}. Bolded items indicate changes from the base macro to implement strategies. In addition to the listed mapping restrictions, all macros are restricted to store different weights in $A_1/A_2$ and different weights in $B_1/B_2$. A flexible model is needed to explore the different components and data movement patterns that are unique to each macro. Open-source models of these macros are available at \href{https://github.com/mit-emze/cimloop}{\(\textit{https://github.com/mit-emze/cimloop}\)}.}
    \label{fig:example_macros}
\pfvs \pfvs \pfvs \pfvs \pfvs \pfvs \end{figure*} 

\begin{itemize}
    \item \textit{Macro A}~\cite{jia} reuses analog outputs across different columns by summing them on wires.
    \item \textit{Macro B}~\cite{sinangil} reuses analog outputs across different columns by summing them with an analog adder.
    \item \textit{Macro C}~\cite{wan} reuses outputs across different cycles by accumulating them with an analog accumulator.
    \item \textit{Macro D}~\cite{wang} uses an analog MAC unit that internally reuses outputs, letting it generate a single output using different weight bits.
    \item \textit{Digital CiM}~\cite{colonnade} reuses outputs digitally to avoid the need for an ADC.
\end{itemize}

Each of these strategies has tradeoffs. Additional components consume area and energy. Components may also introduce mapping and data movement constraints (\eg for \textit{Macro B}, adjacent columns must store different bits of same weight, and for \textit{Macro A}, adjacent columns must store bits from different weights).

In the broad CiM design space, these are just a few sets of possible components, and reducing ADC energy is just one challenge to be addressed.

\textbf{Key modeling challenge 1:} A modeling tool must be \textit{flexible}, meaning able to describe and map workloads to macros with different sets of components and different patterns of data movement/reuse. Furthermore, choices interact with the full system, which may include buffer hierarchies and interconnects. Full-system modeling necessitates a representation that makes it easy to describe complex designs with many different types of components.



To model the design space, it is critical to let users define the components in the system and how they connect. Unfortunately, prior modeling tools do not have this ability. NeuroSim~\cite{neurosim_most_recent} and MNSim~\cite{mnsim_most_recent} model only the \textit{Base Macro} and do not let users add components or define data movement patterns. Timeloop+Accelergy~\cite{timeloop, accelergy, accelergy_pim} let users define architecture-level components (\eg SRAM buffers) but are not able to model circuit data movement or reuse (\eg memory cells and sense amplifiers in an SRAM buffer). For this reason, none of these works can represent $Macros$ $A$, $B$, $C$, or $D$.

\mysubsection{Data-Value-Dependent Energy} \label{slicing_encoding}
Device/circuit energy is \textit{data-value-dependent}, meaning affected by data values propagated by components in the system. We can break data-value-dependence into three components:
\begin{enumerate}
    \item \textbf{DNN Workload:} The system processes DNN operands, which vary between DNN layers/tensors in their distributions, signedness, and amount of sparsity~\cite{raella, sparseloop_1}.
    \item \textbf{Representation:} How data values appear is determined by how the hardware represents operands~\cite{fidelity_encoding_exploration}. First, operands are \textit{encoded}, meaning represented as bits. Next, they are \textit{sliced}, meaning bits are partitioned across different devices and circuits. Representations may change for different tensors and different components in the system.
    \item \textbf{Circuits:} Different circuits spend different amounts of energy to propagate different data values. For example, some ADCs~\cite{low_value_adc, chen2022bit} and DACs~\cite{wang,sinangil} spend less energy to convert small values.
\end{enumerate}

\begin{figure}[]
    \centering
    \svgcap \includegraphics[width=\linewidth, keepaspectratio] {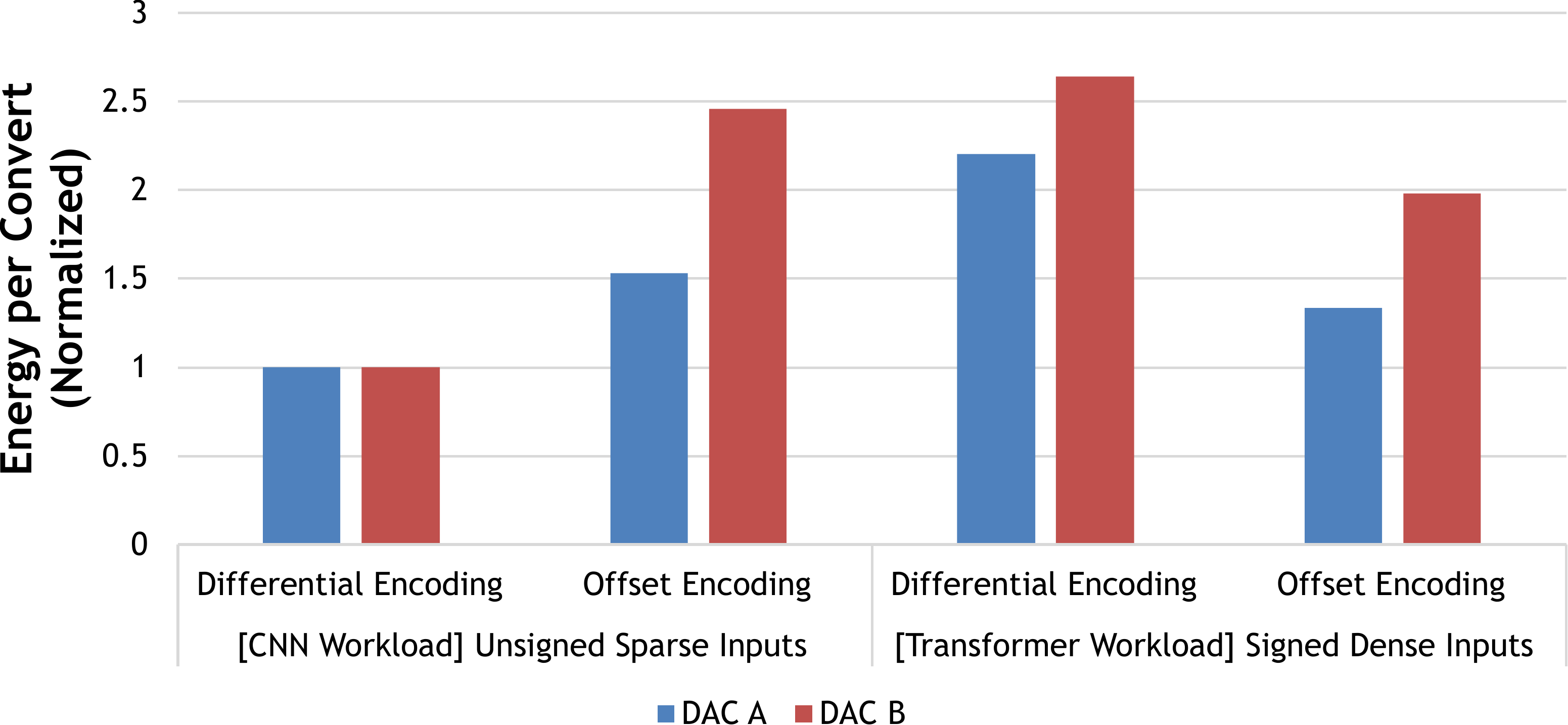}
    \caption{Data-value-dependence can affect circuit energy by ${>2.5\times}$, and its effect is different for each DAC, encoding, and layer. The best encoding is different for each layer.}
    \label{fig:data_value_dependent}
\pfvs \end{figure}

Fig.~\ref{fig:data_value_dependent} shows examples of each of these three components affecting DAC energy. Data-value-dependence can have a $2.5\times$ effect on the energy of the DAC type shown.

\textbf{Key Modeling Challenge 2:} To model accurately, a modeling tool must model the data-value-dependent interactions between workload, data representations, and circuit energy. To explore the space of interactions possible, the tool must do this flexibly (\ie for heterogeneous user-defined circuits).

Unfortunately, Timeloop+Accelergy and MNSim use inaccurate fixed-energy~\cite{accelergy,timeloop} or fixed-power~\cite{mnsim_most_recent} models that do not model data-value-dependence.

\mysubsection{The Need for Fast Modeling}
Co-exploring multiple levels opens a large design space, and the number of design points grows exponentially with the number of decisions explored. Furthermore, for each point, we may map a DNN with hundreds of layers~\cite{gpt2} and search thousands of mappings for each layer~\cite{timeloop}.

\textbf{Key Modeling Challenge 3:} Fast modeling is essential to explore the large design space. This goal is in tension, however, with accurate modeling because accurate modeling requires modeling the many data values that circuits propagate.


Unfortunately, this tension has led prior work to be either accurate or fast, but not both. NeuroSim~\cite{neurosim_most_recent} uses an accurate data-value-dependent model, but it simulates every data value and is therefore slow. Timeloop+Accelergy~\cite{accelergy,timeloop} and MNSim~\cite{mnsim_most_recent} use faster fixed-energy and fixed-power models, but these models do not capture data-value-dependence.


\mysection{CiMLoop}
In this section, we describe CiMLoop and how it addresses the three key challenges described previously. We begin with an overview of the CiMLoop infrastructure. We then describe CiMLoop's flexible per-component data reuse models and CiMLoop's system representation. Following that, we describe CiMLoop's data-value-dependent energy model and provided component models. Finally, we discuss why CiMLoop is fast.

\mysubsection{Infrastructure Overview}
CiMLoop is built on the Timeloop+Accelergy~\cite{accelergy,timeloop} infrastructure. Users describe systems with a collection of YAML files (architecture, workload, components, other configurations) and run the infrastructure with a Python interface. Accelergy performs area/energy estimations for each component in the system. Timeloop uses these estimates to perform mapping (\ie spatial \& temporal tiling) searches and model full-system energy, throughput, and area for systems running DNN workloads. CiMLoop modifies both Accelergy and Timeloop to support CiM features, introducing a component modeling interface that supports data-value-dependent modeling and circuit-level data movement/reuse/mapping support. CiMLoop also introduces a new flexible architecture specification and fast modeling pipeline.


\mysubsection{Flexible Circuit and Architecture Modeling}
CiMLoop's flexible specification lets users specify components, where they are in the system, and how they can move and reuse data. We define \textit{components} as anything that may move or reuse data. Components may be fine-grained (\eg an SRAM bitcell) or coarse-grained (\eg an SRAM buffer).

\subsubsection{Per-Component Data Movement and Reuse}
In a data movement hierarchy, component X \textit{reuses} a piece of data if X uses the data multiple times with one access to the parent of X. Reuse can only occur if reuse is supported by hardware, present in the workload (\ie multiple computations use the same data), and present in the mapping (\ie computations are mapped to adjacent spatial components or timesteps such that reuse can happen between the components or timesteps).

For each component and each tensor (\ie inputs, outputs, weights), users may set supported reuse independently. \textit{Temporal-Reuse} can store data between cycles, \eg buffers temporally reuse data. \textit{No-Temporal-Reuse} does not allow reuse between cycles. No-temporal-reuse components may or may not \textit{coalesce}, meaning change multiple accesses of the same value into one access of backing storage. For example, when an adder sums several values, it coalesces them into one output (\ie reusing the output for multiple additions). On the other hand, a DAC may not coalesce; if the same piece of data is propagated through a DAC multiple times, it must be fetched from backing storage multiple times. Temporal-reuse components can always coalesce if given the opportunity. Additionally, data may \textit{bypass} data around a component without activating that component (\eg inputs bypass a weight buffer). 

Spatially between components, data may be reused (multicast or reduced between components) or not reused (unicast to each component individually). For example, in the \textit{Base} macro topology, outputs are reused between rows (\ie outputs from multiple rows are summed and read once by the ADC) but not columns (\ie outputs from each column are read individually by the ADC). We show each of these reuse options when describing the macro in the following section.

\subsubsection{Representing Systems with Container-Hierarchy} 
CiMLoop uses a container-hierarchy representation to scalably represent circuits and architectures. A \textit{container} is a grouping of components and/or (sub)containers, and a \textit{container-hierarchy} is a series of containers where each contains all subsequent components/containers. Container-hierarchies are useful for describing CiM systems because:

\begin{itemize}
    \item Each container isolates local design decisions while abstracting other containers. This permits easy description of complex designs with many different types of components.
    \item Container hierarchies express circuits and architecture in one hierarchy, so they are compatible with tools and abstractions for memory hierarchies~\cite{timeloop}.
    \item Container-hierarchies can be nested to arbitrary depths (\eg can describe memory cells, circuits, architecture, and multi-chip data movement at the same time).
    \item Multiple container-hierarchies can be mixed and matched to aid design space exploration (\eg a user may create one macro and test that macro in multiple systems).
\end{itemize}

To show how to use container-hierarchy, we use it to describe the CiM system in Fig.~\ref{fig:hierarchy_example}. This system is similar to the macro shown in Fig.~\ref{fig:how_cim_works}, but we add a digital adder to reuse (sum) outputs after the ADC and we use a single buffer as the memory hierarchy. In our description, each \underline{component is underlined} and each \textbf{reuse option is bolded}. The macro can be described as:

\begin{enumerate}
    \item A container abstracts the macro, separating it from the rest of the system. The macro communicates with a \underline{buffer} that \textbf{temporally reuses} inputs and outputs (\ie stores them over time).
    \item In the macro, inputs pass through a \textbf{no-temporal-reuse-no-coalesce} \underline{bank of DACs} to be converted to analog. Meanwhile, a \textbf{no-temporal-reuse} \underline{adder} sums values from columns and \textbf{coalesces} them into one output. Finally, we abstract the array as two column containers. Inputs, but not outputs, are \textbf{reused spatially} between columns (\ie all columns receive the same inputs, but each column produces an independent output).
    \item In each column, two \underline{memory cells} \textbf{temporally reuse} (\ie store over time) weights. Outputs from memory cells are \textbf{reused (summed) spatially} then pass through a \textbf{no-temporal-reuse-no-coalesce} \underline{ADC} to be converted to digital.
\end{enumerate}

\definecolor{container}{rgb}{0.69, 0.0, 0.27}
\definecolor{constraints}{rgb}{0, 0.5, 0}
\def \maroon[#1]{\textcolor{container}{\textbf{#1}}}
\def \green[#1]{\textcolor{constraints}{\textbf{#1}}}

Fig.~\ref{fig:hierarchy_syntax} shows a simplified YAML specification for this CiM macro. Given this specification and a workload, CiMLoop maps (\ie spatially/temporally schedules) the workload onto the system and generates area/energy/throughput estimates. In the full (non-simplified) specification, each component includes a class (\eg a buffer is SRAM), attributes (\eg ADC resolution), and optional constraints/heuristics for the mapping search. CiMLoop tutorials include the full syntax.

\begin{figure}
    \begin{subfigure}{1\linewidth}
    \includegraphics[width=\linewidth, keepaspectratio] {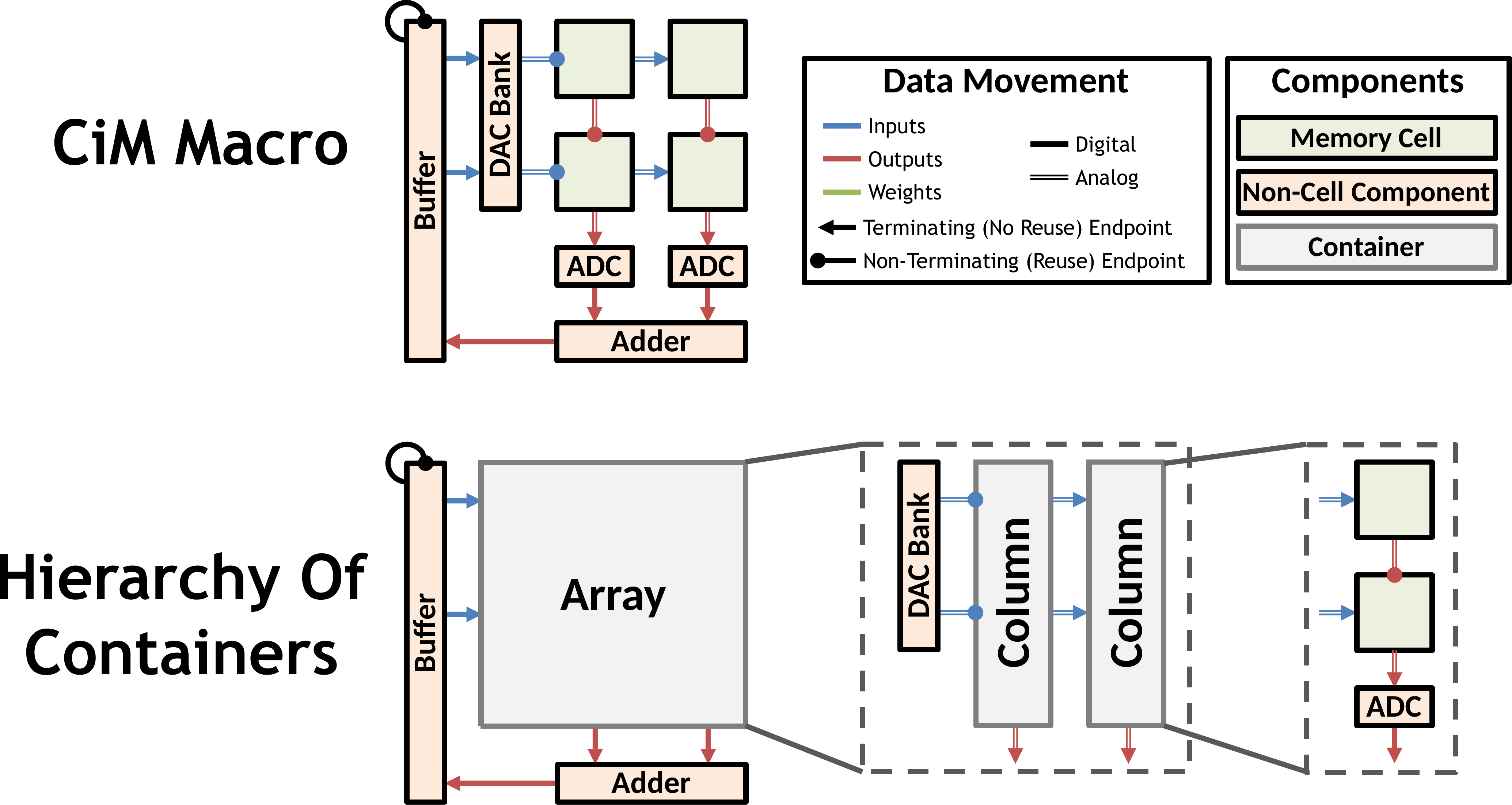}
    \newsubcap{Container-hierarchy partitions systems into simple containers.}\label{fig:hierarchy_example}
    \end{subfigure}

    \begin{subfigure}{1\linewidth}
    \scriptsize \linespread{0.93}
    \begin{minted}[
    escapeinside=||,
    frame=single,
    ]{yaml}
!Component   # Buffer stores inputs & outputs.
name: |\textbf{buffer}|
temporal_reuse: [Inputs, Outputs] # Bypass weights

!Container  # Container includes everything declared in
name: |\textbf{macro}| # following lines

# |\textbf{=================== Now inside Macro ====================}|
!Component  # Adder sums values and coalesces them into
name: |\textbf{adder}| # one output.
coalesce: [Outputs] # Bypasses inputs/weights

!Component # Inputs pass through DACs, convert to analog.
name: |\textbf{\texttt{DAC\_bank}}| # DACs can not coalesce.
no_coalesce: [Inputs] # Bypass outputs/weights

!Container   # Inputs are spatially reused between columns,
name: |\textbf{column}| # while outputs/weights are not.
spatial: {meshX: 2} # 2 columns in X dimension
spatial_reuse: [Inputs] # Reuse inputs, not outputs/weights

# |\textbf{=================== Now inside Column ===================}|
!Component # Outputs pass through ADC, convert to digital
name: |\textbf{ADC}|
no_coalesce: [Outputs] # Bypass inputs/weights

!Component # Memory cells store & temporally reuse weights.
name: |\textbf{\texttt{memory\_cell}}| # Memory cells spatially reuse outputs.
spatial: {meshY: 2} # 2 cells in Y dimension
temporal_reuse: [Weights] # Bypass inputs/outputs
spatial_reuse: [Outputs] # Reuse outputs not inputs/weights

    \end{minted}
    \normalsize
    \centering
    \newsubcap{\label{fig:hierarchy_syntax} A YAML container-hierarchy describes the system in Fig.~\ref{fig:hierarchy_example}. \#-marked comments explain the syntax. \mintinline{yaml}{!Component} / \mintinline{yaml}{!Container} tags declare components/containers, \green[\mintinline{yaml}{temporal_reuse}] / \green[\mintinline{yaml}{coalesce}] / \green[\mintinline{yaml}{no_coalesce}] directives describe data reuse, and \green[\mintinline{yaml}{spatial}] describes the number of each component in the X/Y dimensions. If a particular tensor is not listed for a component, then it bypasses the component.}
    \end{subfigure}
\pfvs \pfvs \end{figure}



\mysubsection{Accurate Data-Value-Dependent Modeling} \label{accurate_modeling}
\subsubsection{Data-Value-Dependent Pipeline} 
To model data-value-dependent energy, CiMLoop needs to know what data values each component propagates and how these impact energy. This procedure is broken into three steps:
\paragraph{Workload Operand Distributions} For each workload tensor (\ie inputs, outputs, weights), CiMLoop uses a distribution of values that operands in this tensor take. Distributions, rather than full tensors, are fast to model as described in Section~\ref{fast_modeling}.

\paragraph{Encoding and Slicing} CiMLoop calculates the representation of elements in each tensor for each component (representations may change as data moves through the system). Representation first depends on \textit{encoding}, meaning representing operands as binary values. CiMLoop supports encoding and slicing functions from CiM implementations, including offset~\cite{isaac}, differential~\cite{raella}, XNOR~\cite{jia}, and magnitude-only~\cite{forms}. Other encoding schemes can be defined or imported. After encoding, binary values are \textit{sliced}, meaning their bits are partitioned across hardware components. Computations across multiple slices are exposed to the Timeloop~\cite{timeloop} mapper, letting CiMLoop tile and spatially/temporally map the bits of each tensor.

\paragraph{Component Energy Modeling} At this point, CiMLoop has the distribution of encoded and sliced data values propagated by each component. Per-component models use these distributions to calculate energy. Each component may use distributions differently (\eg resistor energy increases with the duration of applied voltages, while capacitor energy increases with the amount of switching of applied voltages).

\subsubsection{Provided Models}
CiMLoop includes a suite of provided plug-ins that let users model components in their own systems. CiMLoop uses the Accelergy~\cite{accelergy} plug-in suite, including the CACTI~\cite{CACTI} plug-in modeling buffers and the Aladdin~\cite{aladdin} plug-in modeling digital components. CiMLoop also includes a simple plug-in interface that lets users define new data-value-dependent energy models. To model CiM components, CiMLoop includes the following additional plug-ins.

\textbf{The ADC Plug-In}~\cite{adc_plug_in} uses regression-based models over published ADCs~\cite{adc_survey,ADC_scaling,ADC_Scaling_Murmann,schreier} to predict the area and energy of an ADC (or bank of ADCs) that meets a user-defined throughput, resolution, and number of ADCs.

\textbf{The NeuroSim Plug-In} uses NeuroSim~\cite{neurosim_most_recent} to model array row/column drivers and ADCs; memory cells; and digital components such as adders, multiplexers, and logic gates. To enable flexibility, CiMLoop separates NeuroSim components from one another so they can be reassembled into user-defined systems. To enable fast modeling, CiMLoop connects the NeuroSim plug-in to the fast modeling pipeline described in Section~\ref{fast_modeling}. These changes maintain high accuracy, which we quantify in Section~\ref{evaluation:cimloop_accuracy}. We also connect the NeuroSim plug-in to memory cells in the NVMExplorer~\cite{NVMExplorer} memory cell exploration tool to let users flexibly swap device models.

\textbf{The Library Plug-In} models a library of components used in various CiM works~\cite{aladdin, AtomLayer, BRAHMS, forms, isaac, neurosim_most_recent, PRIME, TIMELY, raella, wan}. The Library plug-in can be used to quickly create new systems by leveraging off-the-shelf component models~\cite{cmos_scaling}, or it can be used to fairly compare different architectures while using a common set of components.

\mysubsection{Fast Modeling} \label{fast_modeling}

To model quickly, CiMLoop calculates the average energy for each action by each type of component. This average energy can then be applied to any number of actions by that type of component, letting CiMLoop model an arbitrary number of components/operations with constant runtime. This is particularly helpful for quickly modeling CiM systems, which may have many components in parallel (\eg thousands of memory cells in an array). 

\subsubsection{Modeling DNN Operand Values with Distributions}
CiMLoop leverages the distributions (\ie a probability mass function for each tensor) of DNN operand values~\cite{raella} as an input to its statistical model. As described in Section~\ref{accurate_modeling}, CiMLoop will use these distributions and their hardware representations to derive the data values that each type of component will propagate.

CiMLoop decouples the gathering of DNN operand distributions from system modeling. This is for two reasons. 

\begin{enumerate}
    \item It allows CiMLoop design space exploration to be much faster, as the DNN does not need to be run to evaluate each CiM design.
    \item It permits multi-fidelity modeling by letting users trade off the fidelity of distributions with the amount of user effort required to obtain distributions. Users may provide CiMLoop with distributions that are \textbf{easy-to-obtain, yet low-fidelity} (\eg model ReRAM device input voltage as a uniform distribution from 0V to 1V); \textbf{moderately-easy-to-obtain and moderate-fidelity} (\eg two's-complement encode a DNN input tensor and use the resulting distribution as ReRAM device input voltages); or \textbf{difficult-to-obtain, yet high-fidelity} (\eg simulate a DAC generating analog voltages and use the resulting distribution as ReRAM device input voltages).
\end{enumerate}


CiMLoop assumes that the distributions of values in separate tensors are independent. Leveraging this assumption, CiMLoop stores an independent distribution for each tensor. Independent distributions, rather than joint distributions, are faster to record and use in energy estimations because for $N$-point probability density functions and $T$ tensors, the number of points that must be stored scales with $O(NT)$ for independent distributions and $O(N^T)$ for joint distributions. We note that this restriction is not fundamental to CiMLoop, and user-defined models may use joint distributions. However, using a joint distribution will make CiMLoop slower, and we found that independent distributions are sufficient to get high accuracy (see Section~\ref{accurate_modeling}).

\subsubsection{Per-Layer Model}
For each DNN layer, plug-in models (described in Section~\ref{accurate_modeling}) for each component receive distributions of data values (one distribution for each tensor) and calculate average per-action energy. Different layers require different per-action energies because data value distributions can change between layers and tensors within layers~\cite{raella}. 

\subsubsection{Mapping-Invariant Energy} \label{mapping_invariant}
CiMLoop assumes that the energy of each action by each component is mapping-invariant (\ie it does not change across different mappings). Note that overall component energy is not mapping-invariant, as the number of actions by each component depends on the mapping (\ie the energy per read of a buffer must be the same across mappings, but the number of reads, and thus overall energy, may be different).

This assumption is valid if the mapping does not affect the distribution of values propagated for any particular tensor. Generally, for regular mappings (\ie those that can be represented by a loop nest~\cite{timeloop,ruby}), this is the case because mappings affect tensor elements equally (\eg if a mapping results in DACs propagating inputs twice as often, then each input element is propagated twice as often and the distribution does not change). This assumption is violated for sparse systems that may skip zero elements~\cite{sparseloop_1, sparseloop_2}, though CiMLoop focuses on dense CiM system modeling.

Based on this assumption, CiMLoop pre-calculates the average energy for each action by each component, amortizing calculation time over many mappings.

\mysubsection{Example Data-Value-Dependent Calculation}
As an example, we will calculate the data-value-dependent energy consumed by ReRAM device reads whose energy $E$ is the product of the weight conductance $G$, the square of the applied input voltage $V$, and the read duration $T_{read}$. 


Algorithm~\ref{alg:reram_power_dse} shows the calculation as it would be performed in a design space exploration over architectures. For each DNN layer (line 2), we profile the DNN to get probability mass functions for input and weight values (line 3). Then, for each architecture in the exploration (line 4), lines 5 and 6 calculate the average squared voltage and average conductance that the architecture will use to represent inputs and weights, respectively. This uses the architecture's slicing and encoding functions as defined in Section~\ref{slicing_encoding} (\eg an architecture may encode an operand as a two's complement value and represent this value as a voltage between 0V and 1V). For simplicity, we do not show slicing (\ie partitioning bits across multiple components and/or timesteps). With slicing, then Algorithm~\ref{alg:reram_power_dse} would find the average power across all slices.

Finally, we calculate average ReRAM read energy in line 7. Next, lines 8 and 9 iterate over different mappings and calculate the number of times each mapping reads ReRAM devices. Line 10 multiplies the number of reads by the average read energy to get total ReRAM read energy.

Amortization of data-value-dependent calculation time is key to CiMLoop's speed. Calculating data-value-dependent energy and profiling DNNs consume negligible overhead because data-value-dependent calculations (lines 5-7) are amortized over the innermost loop (line 8), which may run for thousands of mappings, while profiling the DNN (line 3) is amortized over both the innermost (line 4) and intermediate (line 8) loops. The per-mapping evaluation (line 9), is non-data-value-dependent (based on the assumption in Secton~\ref{mapping_invariant}) and dominates runtime in most explorations.

\begin{algorithm}[ht]
  \small
  \caption{Calculate Data-Value-Dependent ReRAM energy in a design space exploration. Energy $GV^2T_{Read}$ is the product of conductance, squared voltage, and read duration. $P_I(x)$ and $P_W(y)$ are the probability mass functions of inputs and weights.
  $V_I(Arch,x)$ and $G_W(Arch,y)$ encode input and weight values as voltages and conductances, respectively.\label{alg:reram_power_dse}}
  \SetAlgoLined
  \DontPrintSemicolon
  \SetKwProg{Fn}{Func}{}{}
  \SetKwFunction{CalcReRAMEnergy}{CalcReRAMEnergy}
  \Fn{\CalcReRAMEnergy{Architectures, Layers}}{
  \tcc{Requires a set of architectures to explore and a set of DNN layers to run}
    \For{\(Layer \in DNN\ Layers\)} {
        \tcc{Run the layer and record input/weight distributions}
        $P_{I}(x)$, $P_{W}(x)$ = RecordOperandPMFs(Layer) \; 
        \For{\(Arch \in Architectures\)} {
            \tcc{Calculate average data-value- dependent ReRAM read energy}
            $V^2_{Avg}$ = $\sum_{x} P_{I}(x)\times{V_{I}(Arch,x)}^2$ \; 
            $G_{Avg}$ = $\sum_{y} P_{W}(y)\times G_{W}(Arch,y)$ \; 
            $E_{Avg}$ = $V^2_{Avg}\times G_{Avg}\times T_{read}$ \; 
            \For{\(Mapping \in GetMappings(Arch, Layer)\)} {
                \tcc{Mapping evaluation dominates runtime}
                \#Reads = \(Evaluate(Mapping,\ Arch)\) \;
                $Mapping.E_{ReRAM}$ = $E_{Avg} \times \#Reads$\;
          }
      }
    }
}
\end{algorithm}

\mysection{Accuracy and Model Speed Evaluation}
In this section, we evaluate CiMLoop accuracy and speed relative to prior CiM modeling works. We will use NeuroSim~\cite{NeuroSim} as a baseline. Note that CiMLoop uses the NeuroSim plug-in, so any accuracy or speed differences are due to CiMLoop's data-value-dependent and fast modeling pipelines. We test using ResNet18~\cite{ResNet} and ImageNet~\cite{imagenet}. Other NeuroSim-provided models use the CIFAR-10~\cite{cifar10} dataset which is less applicable to modern large-scale DNNs.

\mysubsection{CiMLoop Accuracy}~\label{evaluation:cimloop_accuracy}
To evaluate the accuracy of CiMLoop's statistical data-value-dependent model, we model the NeuroSim~\cite{neurosim_most_recent} macro with default parameters. NeuroSim, which calculates every data value propagated by every modeled component, is used as a ground truth. We compare the following:
\begin{itemize}
    \item \textbf{CiMLoop} uses an input, output, and weight distribution for each DNN layer as described in Section~\ref{accurate_modeling}.
    \item \textbf{Non-Data-Value-Dependent} is a fixed-energy model~\cite{accelergy}. We optimistically assume that energy is calculated using data values averaged over all layers. In general, the fixed-energy model would not incorporate any knowledge of the DNN.
\end{itemize}

\begin{figure}[]
    \centering
    \svgcap \includegraphics[width=\linewidth, keepaspectratio] {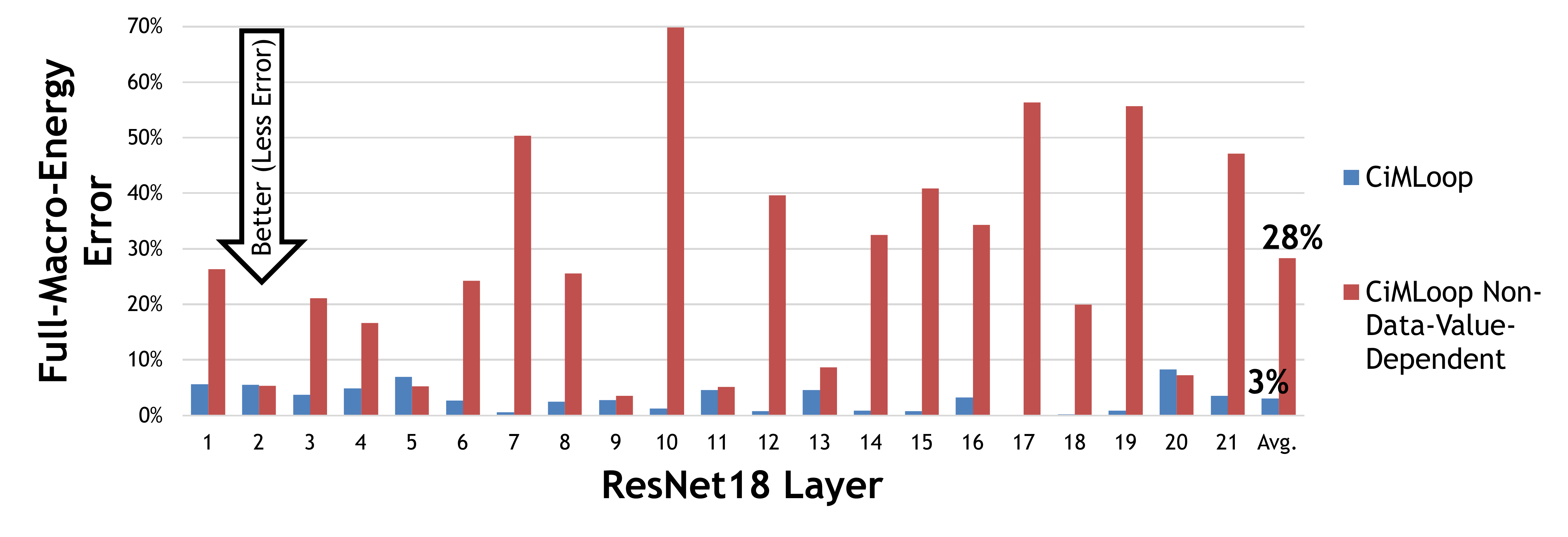}
    \caption{CiMLoop's data-value-dependent model is significantly more accurate than non-data-value-dependent models.}
    \label{fig:accuracy}
\end{figure}

Fig.~\ref{fig:accuracy} compares the accuracy of these two setups when measuring full-macro energy across different layers of ResNet18~\cite{ResNet}. Relative to the NeuroSim ground-truth estimates, CiMloop's data-value-dependent model achieves an average/max error of 3\%/7\% while the fixed-energy model has an average/max error of 28\%/70\%. The fixed-energy model has high error because it does not model the distributions of inputs, outputs, and weights, which vary across DNN layers and significantly affect analog component energy. CiMLoop's data-value-dependent model accounts for these distributions, reducing error. CiMLoop's remaining error is due to statistical model error from representing the values of tensors in each DNN layer using independent distributions.

\mysubsection{CiMLoop Speed}~\label{evaluation:cimloop_speed}
We compare the modeling speed of CiMLoop and NeuroSim when running the default NeuroSim macro running ResNet18/ImageNet~\cite{ResNet,imagenet} one input image on an Intel Xeon Gold 6444Y processor. For CiMLoop we show both single-core and 16-core performance running one or 5000 mappings. NeuroSim does not support multithreading or mapping exploration at the time of testing, so it is tested with one core and one mapping.

Table~\ref{tab:speedup} shows that CiMLoop improves modeling speed over NeuroSim by several orders of magnitude. Time per mapping decreases for many mappings because CiMLoop can amortize one-time startup costs of library invocation, intermediate file generation, and data-value-dependent energy modeling. In most explorations, we test thousands of mappings per system and startup consumes negligible runtime.

\begin{table}
\centering \small
\begin{tabular}{cccc}
\textbf{Model} & \textbf{\# Cores} & \multicolumn{2}{c}{\textbf{(Mappings$\times$Layers) / Second}} \\
&& 1 Mapping & 5000 Mappings \\
\midrule
NeuroSim~\cite{neurosim_most_recent} & 1 & 0.07 & - \\
CiMLoop & 1 & 0.28 & 83 \\
CiMLoop & 16 & 2.25 & 1076 \\
\end{tabular}
\caption{CiMLoop is orders-of-magnitude faster than prior accurate modeling works. CiMLoop is faster for more mappings because it amortizes mapping-invariant calculation and up-front invocation time.}
\label{tab:speedup}
\end{table}

CiMLoop's speed benefits would increase with larger systems and/or larger DNN workload tensors because NeuroSim would simulate the additional components and/or operations, while the runtime of CiMLoop's statistical model would not increase (see Section~\ref{fast_modeling}). CiMLoop also optimizes Timeloop+Accelergy~\cite{accelergy,timeloop} modeling speed, making CiMLoop the fastest modeling tool in Table~\ref{tab:prior_works}.

\mysection{Case Studies: Published CiM Macros}~\label{case_studies}
To validate CiMLoop and provide examples of some of the decisions that CiMLoop can explore, we present models of four recently-published fabricated CiM macros~\cite{jia,sinangil,wan,wan_ii,wang,wang_ii}. Each macro includes multiple contributions, but due to space limitations, we only discuss those necessary to validate CiMLoop's results. We encourage the reader to see the authors' publications and our open-sourced models for each macro.

For each macro, we model the array and row/column drivers using the NeuroSim~\cite{neurosim_most_recent} plug-in. We create memory cell models and calibrate the area/energy of each component to match published values. Unless otherwise stated, all dataflows are weight-stationary with weights pre-loaded into the macro's CiM array, inputs sent to the CiM array, and outputs read from the CiM array. Table~\ref{tab:validation} shows parameterized attributes of each of the CiM macros. CiMLoop can model systems with varied technology nodes, bit precisions, and device technologies. For inputs to DNN workloads, we use ImageNet~\cite{imagenet} inputs and Wikipedia~\cite{wiki:Bread} for image and language models, respectively.

\begin{table}
\vspace{2mm}

\addtolength{\tabcolsep}{-0.5em}
\centering \small
\begin{tabular}{lcccccc}
\textbf{Macro} & \shs{Node\\(nm)} & Device & \shs{Input\\Bits} & \shs{Weight\\Bits} & \shs{Array\\Rows$\times$Cols} & \shs{ADC\\Bits} \\
\midrule
\textit{A}\cite{jia}          & 65  & SRAM  & 1-8 & 1-8    & 768$\times$768 & 8    \\
\textit{B}\cite{sinangil}     & 7   & SRAM  & 4   & 4      & 64$\times$64   & 4    \\
\textit{C}\cite{wan,wan_ii}   & 130 & ReRAM & 1-8 & Analog & 256$\times$256 & 1-10 \\
\textit{D}\cite{wang,wang_ii} & 22  & SRAM  & 8   & 8      & 512$\times$128*& 8    \\
\cmidrule{2-7}
& \multicolumn{6}{r}{*Activates a 64$\times$128 subset of the array at once.}
\end{tabular}
\caption{Parameterized attributes of \textit{Macros A-D}.}
\label{tab:validation}
\end{table}

\mysubsection{Validating CiMLoop}
In this section, we validate CiMLoop models against published \textit{Macros A-D}. We compare CiMLoop-modeled results with the published data for each macro, which include both simulated and silicon-measured results. For each of the following plots, we include macros for which we could find published data (\eg the papers for \textit{Macros A/B/D}, but not \textit{C}, have voltage sweep results). For ease of visualization, plots show a representative subset of validation data. We report average percent error measurements using all validation data, including data that we do not show in plots.

\subsubsection{Energy/Throughput and Supply Voltage} 
\begin{figure}[]
    \centering
    \svgcap \includegraphics[width=\linewidth, keepaspectratio] {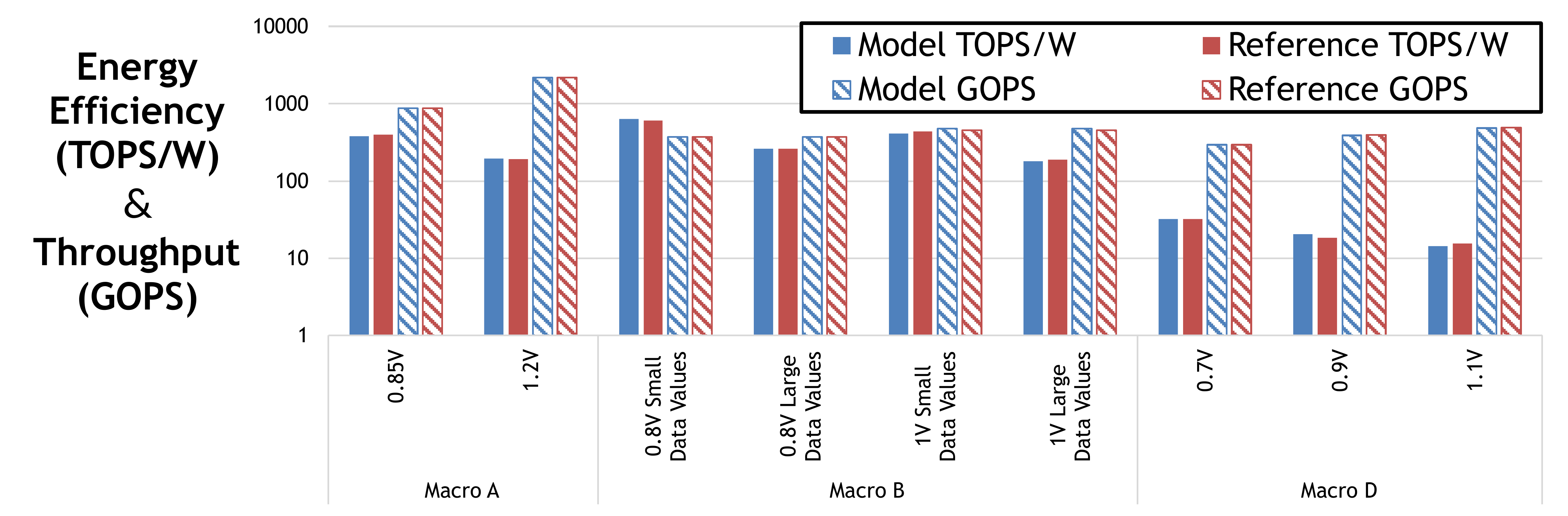}
    \caption{\label{fig:val_vdd} Validating energy/throughput for varied supply voltage.}
\pfvs \end{figure}

Fig.~\ref{fig:val_vdd} validates CiMLoop modeling \textit{Macros A/B/D} energy and throughput across varying supply voltages. \textit{Macro B} energy is data-value-dependent, so we show results for small and large data values. For supply voltage sweeps, CiMLoop's average energy efficiency and throughput errors are 7\% and 2\%.

\subsubsection{Energy/Throughput and Input Encoding}
\begin{figure}[]
    \centering
    \svgcap \includegraphics[width=\linewidth, keepaspectratio] {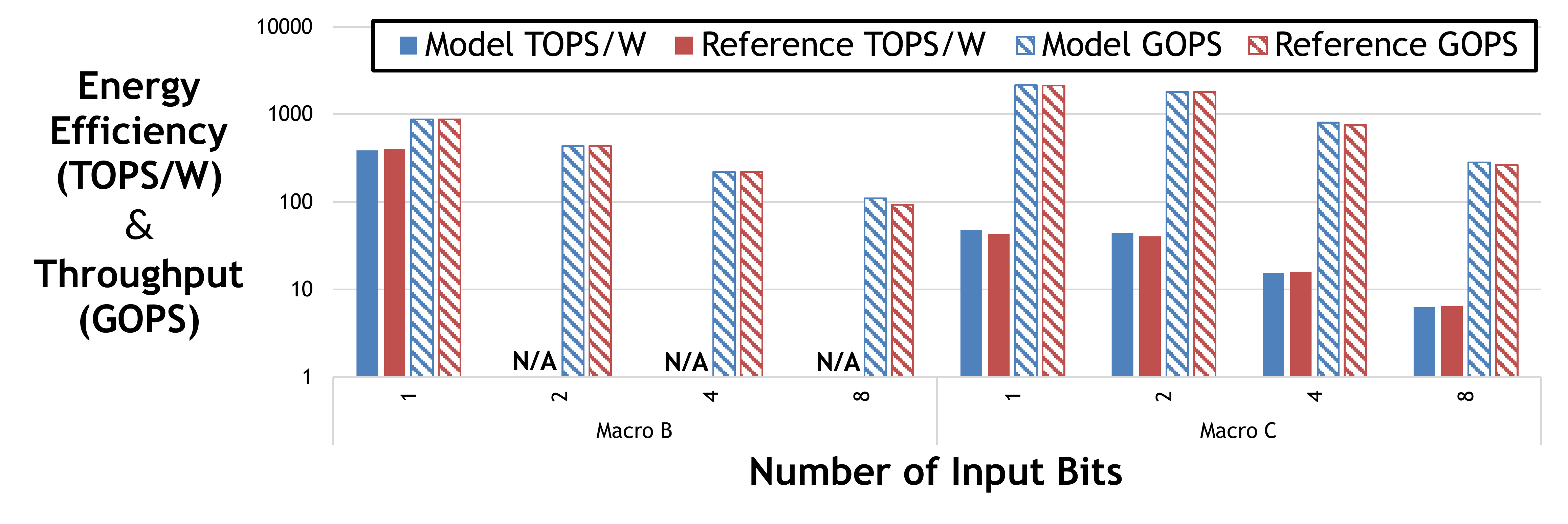}
    \caption{\label{fig:val_bits} Validating energy/throughput for varied \# of input bits.}
\pfvs \end{figure}

Fig.~\ref{fig:val_bits} validates CiMLoop modeling \textit{Macros B/C} energy and throughput for varying numbers of input bits. For these sweeps, CiMLoop's energy efficiency and throughput errors are 6\% and 5\%.

\subsubsection{Energy Breakdown}
\begin{figure}[]
    \centering
    \svgcap \includegraphics[width=\linewidth, keepaspectratio] {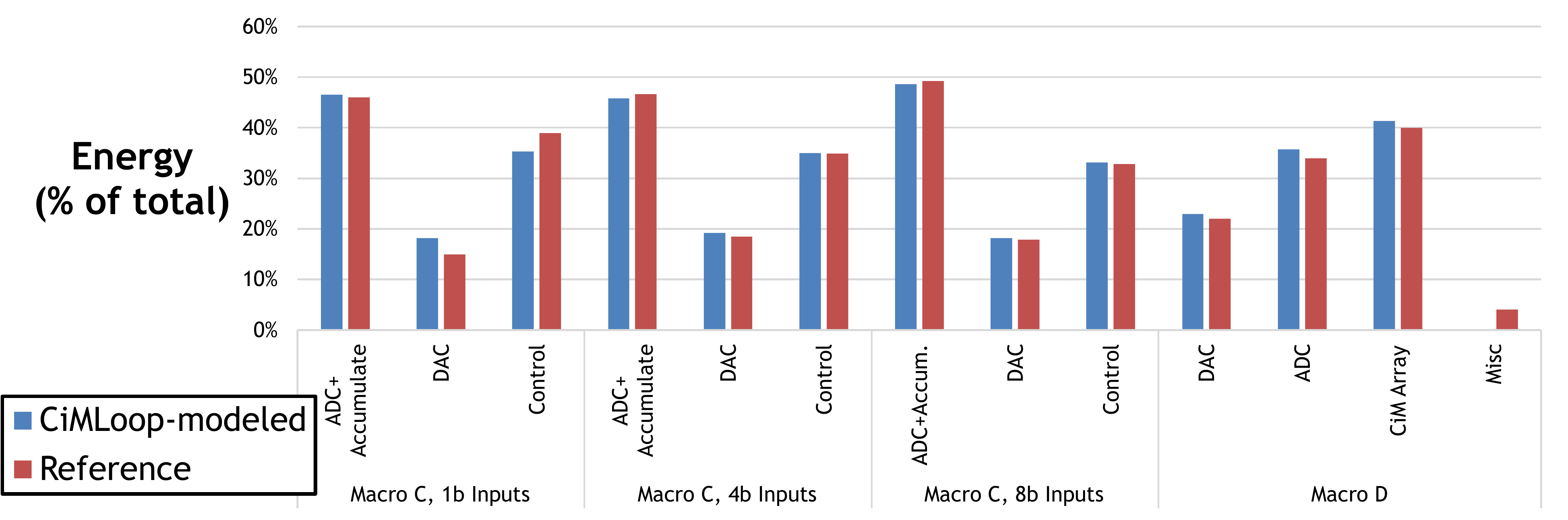}
    \caption{\label{fig:val_energy_breakdown} Validating CiMLoop-modeled energy breakdown.}
\pfvs \end{figure}

Fig.~\ref{fig:val_energy_breakdown} validates CiMLoop modeling the energy breakdowns of \textit{Macros C and D}. For \textit{Macro C}, we report energy for 1b, 2b, and 8b inputs to show that CiMLoop can capture how the energy of each component scales with the number of input bits. Modeled \textit{Macro D} miscellaneous energy is lower than the reference due to components we did not model; accuracy could be improved by adding more components into our model (recall that models are user-defined and components can be added easily). For discrete components, CiMLoop's average energy error is 4\%.

\subsubsection{Area Breakdown}
\begin{figure}[]
    \centering
    \svgcap \includegraphics[width=\linewidth, keepaspectratio] {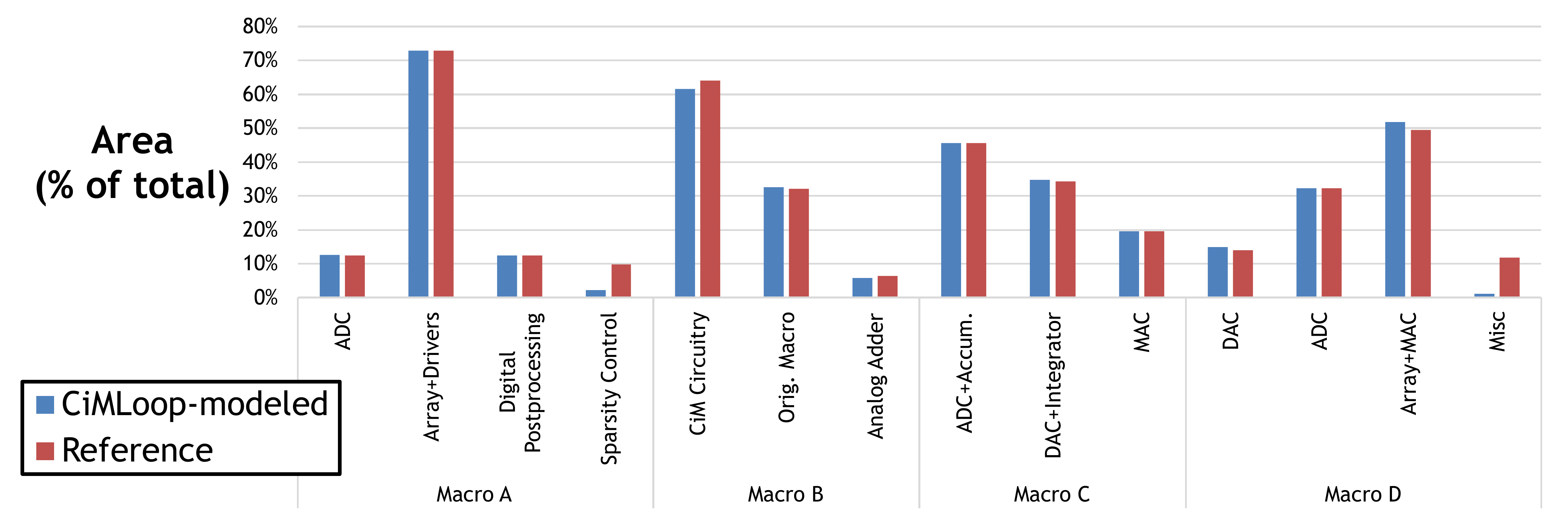}
    \caption{\label{fig:val_area_breakdown} Validating CiMLoop-modeled area breakdown.}
\pfvs \end{figure}

Fig.~\ref{fig:val_area_breakdown} validates CiMLoop modeling the area breakdowns of \textit{Macros A/B/C/D}. As with the energy breakdown, modeled \textit{Macro D} miscellaneous area could be made more accurate by modeling miscellaneous components. For discrete components, CiMLoop's average area error is 8\%.

\subsubsection{Data-Value-Dependent Energy}
\begin{figure}[]
    \centering
    \svgcap \includegraphics[width=\linewidth, keepaspectratio] {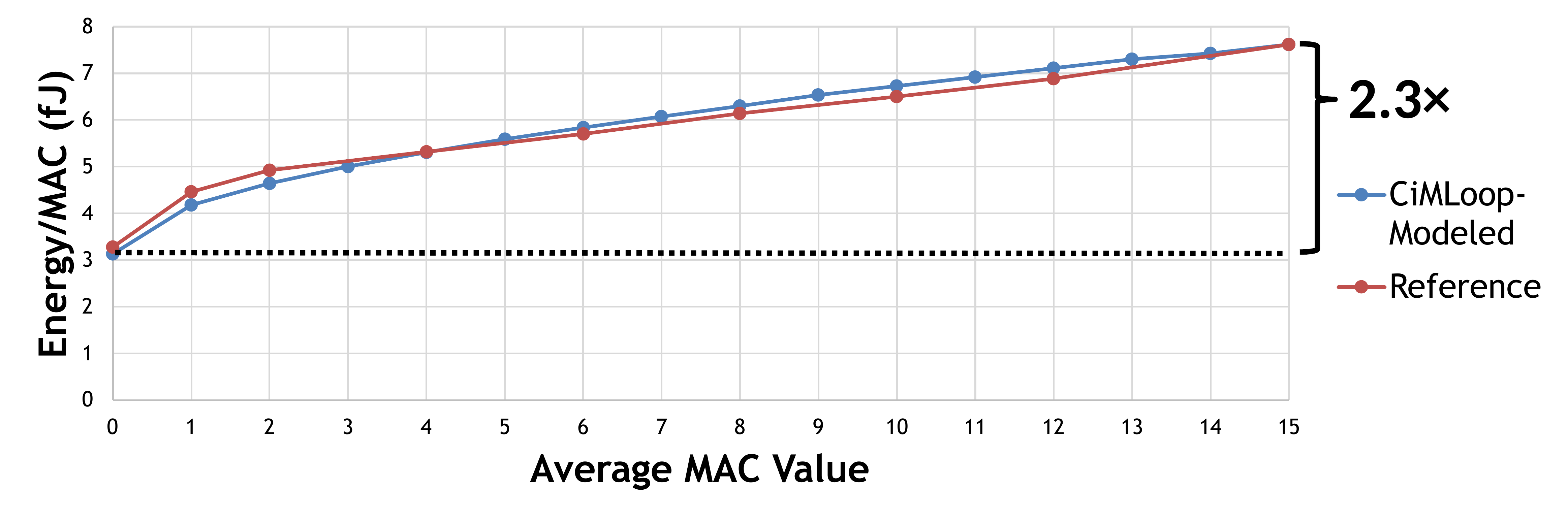}
    \caption{\label{fig:val_data_value_dependent_energy} Validating CiMLoop data-value-dependent energy.}
\pfvs \end{figure}

Fig.~\ref{fig:val_data_value_dependent_energy} validates CiMLoop modeling the data-value-dependent energy of \textit{Macro B}. As average MAC value increases, \textit{Macro B}'s DAC switches more often to supply larger input values and its analog adder charges/discharges with larger analog values. These data-value-dependent effects can increase macro energy by 2.3$\times$.

\mysubsection{Exploring with CiMLoop}
In this section, we show the usefulness of CiMLoop. We explore one level of the CiM stack using each of \textit{Macros A-C} and model a full system using \textit{Macro D}. Finally, we perform a cross-macro comparison.

\subsubsection{Mapping} \textit{Macro A}~\cite{jia} reuses outputs, rather than inputs, between adjacent CiM array columns (Shown in Fig.~\ref{fig:example_macros}). Reusing outputs, and not inputs, between every $N$ array columns increases output reuse $N\times$ but decreases input reuse $N\times$. This affects DAC/ADC energy; more output reuse can decrease ADC converts and ADC energy, but trading off input reuse increases DAC converts and DAC energy. This decision also changes available mappings; if adjacent array columns reuse outputs, then we must map workload operations to share outputs between those columns (\ie map one output channel to those columns). Otherwise, the array is underutilized.

\begin{figure}[]
    \centering
    \svgcap \includegraphics[width=\linewidth, keepaspectratio] {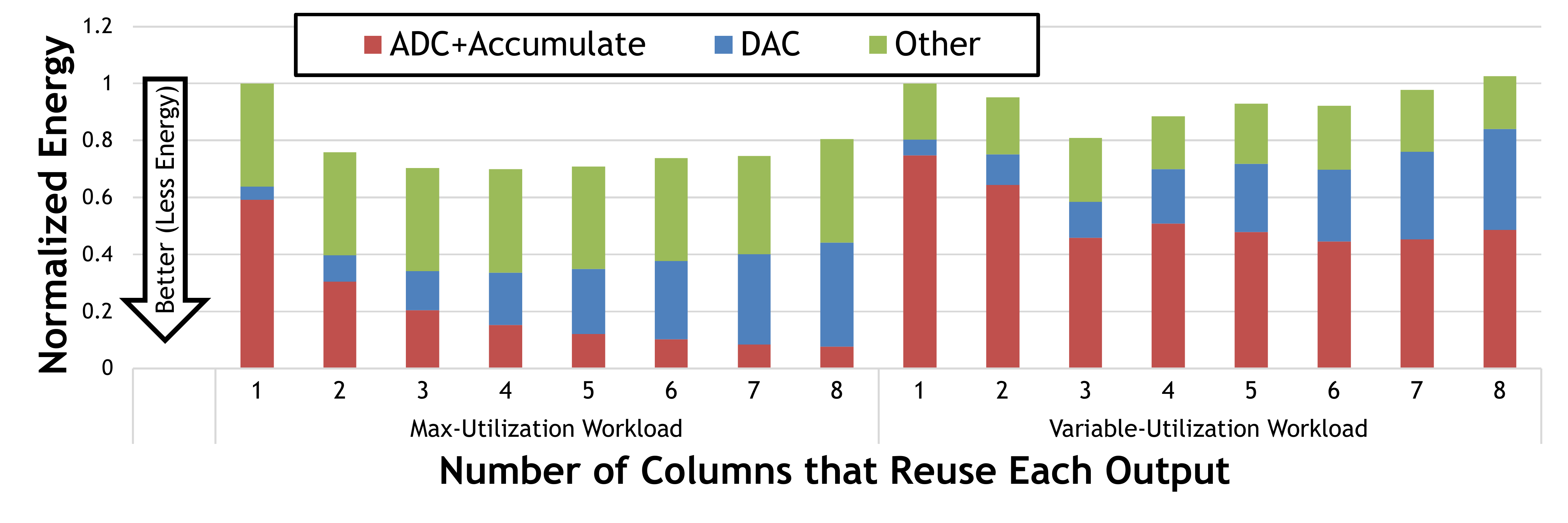}
    \caption{\label{fig:jia_explore} \textbf{\textit{Macro A} + Mapping}: Output reuse between columns decreases ADC energy but trades off lower input reuse and higher DAC energy. The three-column-reuse case had the best-utilization mappings for the variable-utilization workload.}
\pfvs \end{figure}

In Fig.~\ref{fig:jia_explore} we explore \textit{Macro A} configurations that reuse outputs between different numbers of columns. We test configurations when running a maximum-utilization (matrix-vector multiply with dimensions matching the array) and variable-utilization (ResNet18~\cite{ResNet}) workloads. We report DAC, ADC, and other energy. For the maximum-utilization workload, increasing output reuse reduces lower ADC energy but trades off higher DAC energy. For the variable-utilization workload, the three-column-reuse configuration is uniquely lower energy than the other configurations. This is because ResNet18~\cite{ResNet} uses many $3\times3$ convolutional kernels, and they were able to achieve high-utilization mappings on the three-column-reuse macro. This is one of the reasons why Jia et al. use three-column-reuse in their fabricated chip~\cite{jia}.

\subsubsection{Circuits} 
\begin{figure}[]
    \svgcap \includegraphics[width=\linewidth, keepaspectratio] {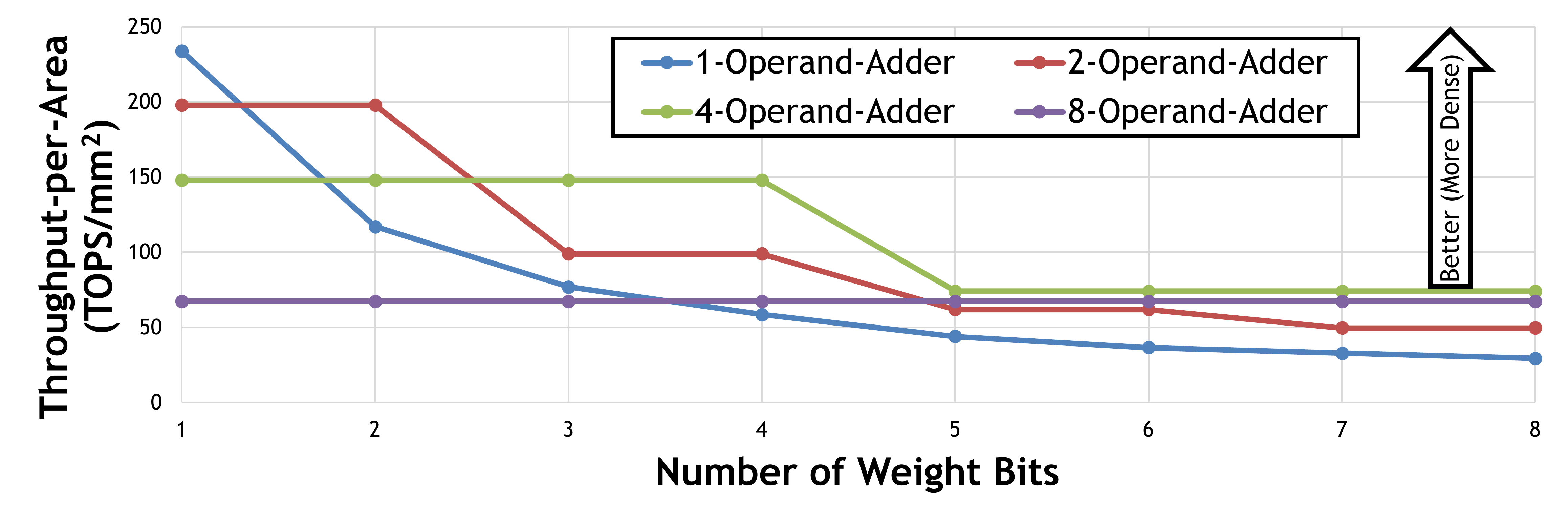}
    \caption{\label{fig:sinangil_explore} \textbf{\textit{Macro B} + Circuits}: An analog adder trades off flexibility for compute density. Fewer-column adders can flexibly leverage fewer-bit weights, but more-column adders achieve higher compute density with more-bit weights.}
\pfvs \end{figure}

\textit{Macro B}~\cite{sinangil} uses an analog adder circuit that sums analog outputs. The ADC reads the resulting sum, rather than each output individually, so the adder can decrease the number of times the ADC is used. This adder comes with the constraint that its inputs must come from different bits of same weights. In Fig.~\ref{fig:sinangil_explore} we explore different widths of analog adders (1-, 2-, 4-, and 8-operand) and workloads with different numbers of bits per weight. Adders that sum more analog operands can increase throughput-per-area because they reduce the number of ADCs required to read outputs. However, they become underutilized when there are fewer bits per weight. Larger adders also consume more chip area; for this reason, the macro with the 8-operand adder never has the highest throughput-per-area.


\subsubsection{Architecture} Using \textit{Macro C}~\cite{wan, wan_ii}, we conduct an architectural exploration over varying array sizes. We set the number of array rows and columns to 64, 128, 256, 512, or 1024 and test maximum-utilization (matrix-vector multiplication), large-tensor-size (Vision Transformer ViT~\cite{vit}), medium-tensor-size (ResNet18~\cite{ResNet}), and small-tensor-size (MobileNetV3~\cite{mobilenetv3}) workloads.

\begin{figure}[]
    \centering
    \svgcap \includegraphics[width=\linewidth, keepaspectratio] {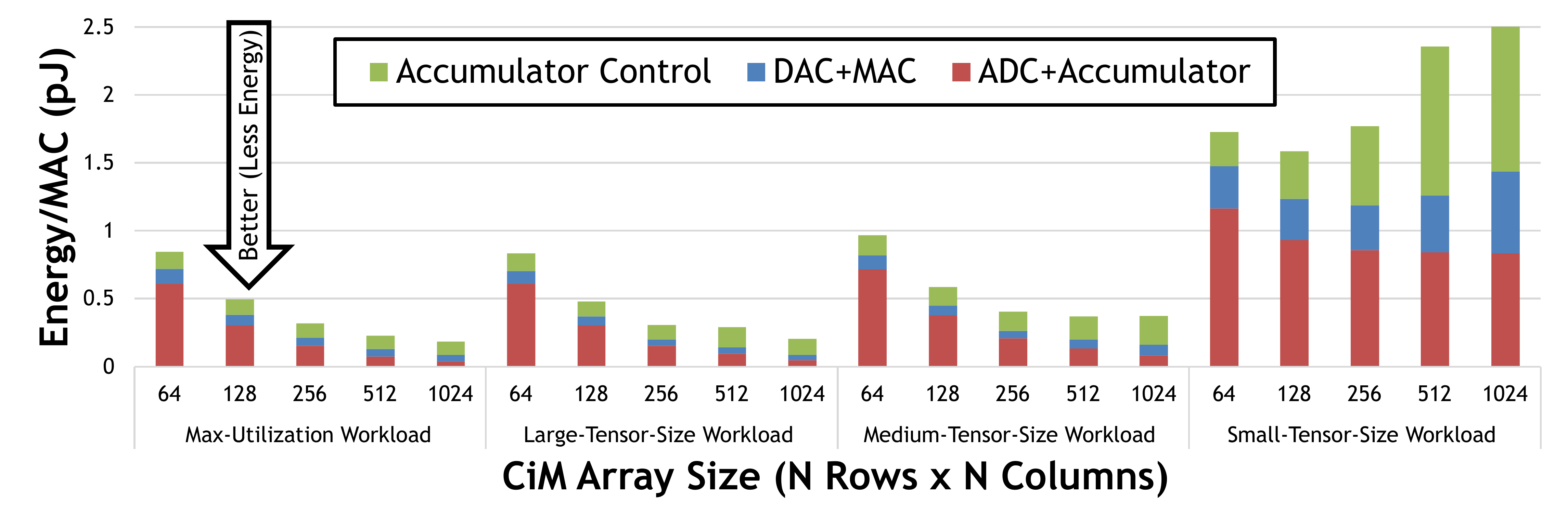}
    \caption{\label{fig:wan_explore} \textbf{\textit{Macro C} + Architecture}: Larger arrays can reduce energy if workload tensors are large enough to utilize them.}
\pfvs \end{figure}

Fig.~\ref{fig:wan_explore} shows that as array size increases, energy decreases due to additional MACs amortizing ADC and digital output sum energy. These effects are strongest for the maximum-utilization and large-tensor-size workloads. For the medium-tensor-size workload, effects saturate as the array grows larger and becomes underutilized for smaller layers. For the small-tensor-size workload, underutilization increased energy for all array sizes and a smaller array was the lowest-energy choice.

\subsubsection{Full-System} CiM systems reduce energy by reducing off-chip data movement and by using CiM arrays to execute many MACs in parallel. To explore the full-system effects of these contributions, we put \textit{Macro D}~\cite{wang, wang_ii} in a full system. The system includes a DRAM backing storage~\cite{CACTI} and a chip that has parallel macros with input/output buffers, routers~\cite{isaac}, and a global buffer. The on-chip memory can fit any tested layer, so inputs/outputs/weights will be transferred no more than once to/from DRAM for each layer. 

Tensor size influences the number of MACs performed, so we test large-tensor (large language model GPT-2~\cite{gpt2}) and mixed-size-tensor (ResNet18~\cite{ResNet}) workloads. We compare three scenarios: \textbf{(1)} inputs/outputs/weights stored off-chip in DRAM and fetched for each layer; \textbf{(2)} inputs/outputs fetched from DRAM, weights stationary (pre-loaded for each layer); and \textbf{(3)} weights stationary, inputs/outputs kept on-chip in the global buffer between layers~\cite{fusion_i, fusion_ii}.

\begin{figure}[]
    \centering
    \svgcap \includegraphics[width=\linewidth, keepaspectratio] {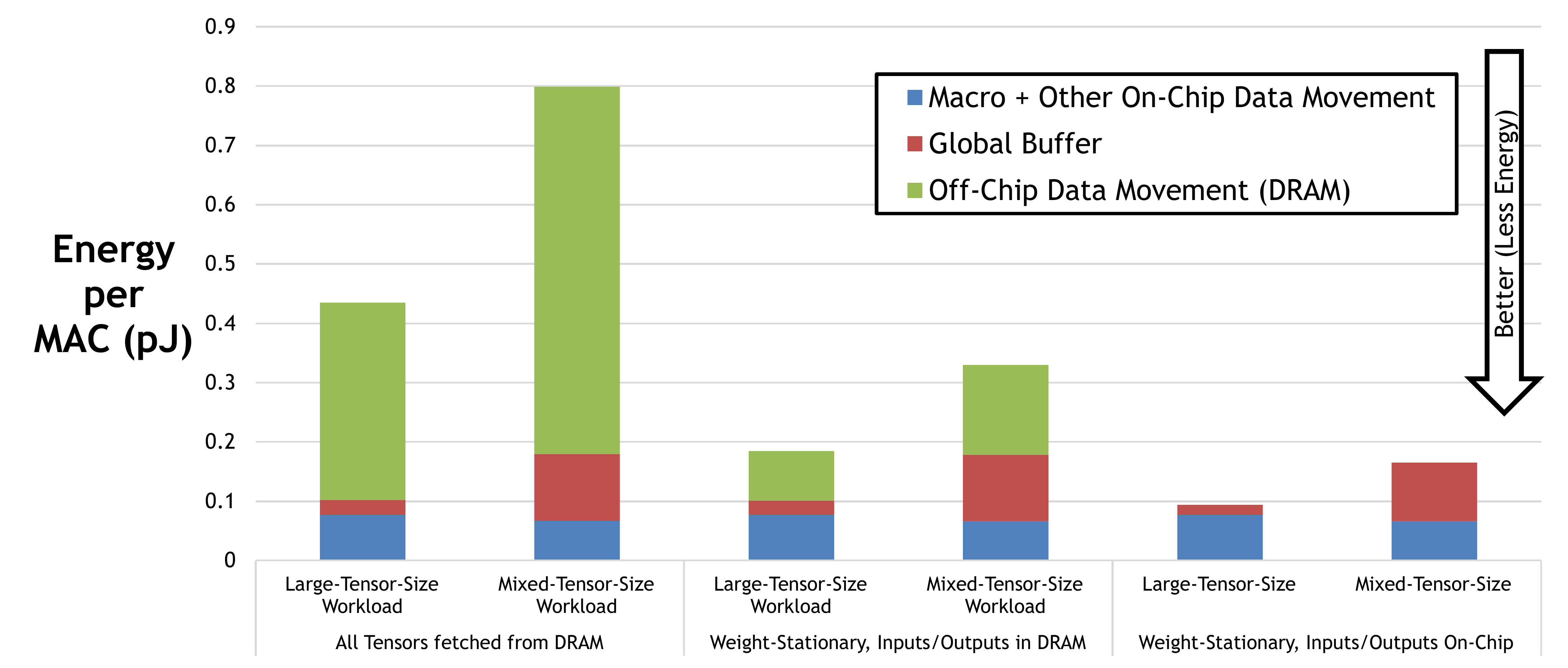}
    \caption{\textbf{\textit{Macro D} + Full-System}: \label{fig:wang_explore} Weight-stationary CiM saves significant energy, but benefits are limited by off-chip movement of DNN inputs/outputs between each layer. CiM will be most effective when combined with layer fusion to keep all tensors on-chip.}
\pfvs \end{figure}

Fig.~\ref{fig:wang_explore} shows the system energy breakdown for on-chip data movement, on-chip global buffer, and off-chip data movement. We see a significant reduction in overall energy when going from off-chip weights to weight-stationary due to fewer weight fetches. Benefits are limited, however, by input/output fetch from off-chip. Recall that this system only transfers inputs/outputs to/from DRAM once per layer; to see further benefits, it is necessary to avoid transferring inputs/outputs off-chip between layers~\cite{fusion_i,fusion_ii}. We note that weight-stationary CiM requires sufficient memory to keep all DNN weights on-chip. To store large DNNs, this may require a multi-chip pipeline~\cite{isaac, PipeLayer} or dense storage technologies~\cite{NVMExplorer}. 

\subsubsection{Cross-Macro} CiMLoop can be used to project how a macro will scale to a new technology node and to fairly compare across different macros. We compare the three SRAM-based \textit{Macros A/B/D}, scaling all macros to 7nm, using \textit{Macro B} memory cells and using an 8b ADC.

\begin{figure}[]
    \centering
    \svgcap \includegraphics[width=\linewidth, keepaspectratio] {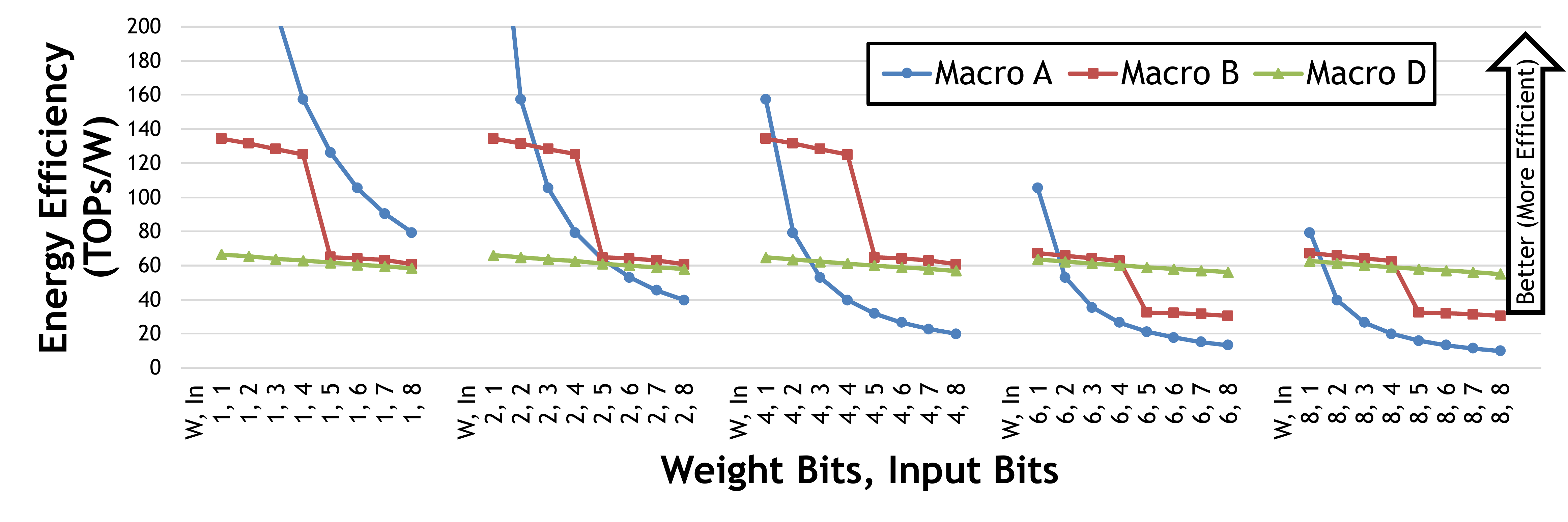}
    \caption{\label{fig:cross_macro} \textbf{\textit{Macros A, B, D} + Cross-Macro}: CiMLoop can fairly compare CiM implementations. The lowest-energy macro choice depends on input/weight bit precisions.}
\pfvs \end{figure}

Fig.~\ref{fig:cross_macro} compares the energy efficiency of all three macros for different numbers of input and weight bits. \textit{Macro A} computes analog MACs with 1b inputs/weights and accumulates results digitally. This lets \textit{Macro A} flexibly leverage few-bit inputs/weights to increase energy efficiency, but the strategy is less efficient with more-bit inputs/weights. \textit{Macros B/D} use 4b/8b analog components (shown in Fig.~\ref{fig:example_macros}) that can increase output reuse and reduce ADC energy. However, these components are underutilized with few-bit inputs/weights, so \textit{Macros B/D} gain little energy efficiency from few-bit operands.

\mysection{Related Works}
In addition to the discussed works~\cite{timeloop, accelergy, accelergy_pim, sparseloop_1, sparseloop_2,  mnsim_1, mnsim_most_recent, neurosim_1, neurosim_2, neurosim_3, neurosim_most_recent}, many works have explored different parts of CiM. While CiMLoop models area/energy/throughput, IBM AI Hardware Kit~\cite{aihwkit}, CrossSim~\cite{CrossSim}, and MemTorch~\cite{memtorch}, model DNN accuracy. Eva-CiM~\cite{eva-cim} models the CPU/CiM interface but not CiM macros. Simeuro~\cite{simeuro} and SuperNeuro~\cite{superneuro} model spiking (rather than deep) neural network systems. PUMA~\cite{puma} provides a detailed model of a particular DNN system but does not explore the design space. Sparseloop~\cite{sparseloop_1,sparseloop_2}, like CiMLoop, uses statistical analytical models, applying them to model sparse DNN~\cite{sparse_dnn} systems.

Sun et al.~\cite{sun2023analog} is a contemporaneous work that combines a parameterizable CiM macro model with a flexible architectural specification~\cite{mei2021zigzag}. Non-parameterizable changes (\eg adding or removing components, changing connections between components) are done by modifying the simulator source code. This contrasts with CiMLoop, which supports non-parameterizable changes by only changing the input specification rather than the simulator source code.

\mysection{Conclusion}
In this paper, we presented CiMLoop: a flexible, accurate, and fast model that connects all levels of the CiM stack. CiMLoop lets researchers evaluate design decisions at each level, co-design across levels, and fairly compare CiM implementations. By bringing all levels together in one model, CiMLoop can bridge the device, circuits, and architecture research communities. CiMLoop can even be used beyond CiM to model traditional~\cite{timeloop} accelerators and those that use other paradigms such as photonics~\cite{photonics}. We hope that researchers will use CiMLoop to share their work, publish open-source models, and reveal new insights and co-design opportunities that leverage the contributions of all communities.

\mysection{Acknowledgments}
We thank the authors of the presented macros Hongyang Jia, Hyunjoon Kim, Mahmut Sinangil, Weier Wan, and Hechen Wang for providing invaluable feedback and advice as we modeled their works. We also thank Anni Lu for her guidance as we integrated NeuroSim as a plug-in to CiMLoop. This work was funded in part by Ericsson, TSMC, the MIT AI Hardware Program, and MIT Quest.

\bibliographystyle{IEEEtran}
\bibliography{main.bib}

\begin{thebibliography}{10}
\providecommand{\url}[1]{#1}
\csname url@samestyle\endcsname
\providecommand{\newblock}{\relax}
\providecommand{\bibinfo}[2]{#2}
\providecommand{\BIBentrySTDinterwordspacing}{\spaceskip=0pt\relax}
\providecommand{\BIBentryALTinterwordstretchfactor}{4}
\providecommand{\BIBentryALTinterwordspacing}{\spaceskip=\fontdimen2\font plus
\BIBentryALTinterwordstretchfactor\fontdimen3\font minus \fontdimen4\font\relax}
\providecommand{\BIBforeignlanguage}[2]{{%
\expandafter\ifx\csname l@#1\endcsname\relax
\typeout{** WARNING: IEEEtran.bst: No hyphenation pattern has been}%
\typeout{** loaded for the language `#1'. Using the pattern for}%
\typeout{** the default language instead.}%
\else
\language=\csname l@#1\endcsname
\fi
#2}}
\providecommand{\BIBdecl}{\relax}
\BIBdecl

\bibitem{NVMExplorer}
L.~Pentecost, A.~Hankin, M.~Donato, M.~Hempstead, G.-Y. Wei, and D.~Brooks, ``{NVMExplorer}: A framework for cross-stack comparisons of embedded non-volatile memories,'' in \emph{2022 IEEE International Symposium on High-Performance Computer Architecture (HPCA)}, 2022, pp. 938--956.

\bibitem{isaac}
A.~Shafiee, A.~Nag, N.~Muralimanohar, R.~Balasubramonian, J.~P. Strachan, M.~Hu, R.~S. Williams, and V.~Srikumar, ``{ISAAC}: A convolutional neural network accelerator with in-situ analog arithmetic in crossbars,'' in \emph{2016 ACM/IEEE 43rd Annual International Symposium on Computer Architecture (ISCA)}, 2016, pp. 14--26.

\bibitem{neurosim_1}
X.~Peng, S.~Huang, Y.~Luo, X.~Sun, and S.~Yu, ``{DNN+NeuroSim}: An end-to-end benchmarking framework for compute-in-memory accelerators with versatile device technologies,'' in \emph{2019 IEEE International Electron Devices Meeting (IEDM)}, 2019, pp. 32.5.1--32.5.4.

\bibitem{neurosim_2}
P.-Y. Chen, X.~Peng, and S.~Yu, ``{NeuroSim}+: An integrated device-to-algorithm framework for benchmarking synaptic devices and array architectures,'' in \emph{2017 IEEE International Electron Devices Meeting (IEDM)}, 2017, pp. 6.1.1--6.1.4.

\bibitem{neurosim_3}
A.~Lu, X.~Peng, W.~Li, H.~Jiang, and S.~Yu, ``{NeuroSim} validation with 40nm {RRAM} compute-in-memory macro,'' in \emph{2021 IEEE 3rd International Conference on Artificial Intelligence Circuits and Systems (AICAS)}, 2021, pp. 1--4.

\bibitem{neurosim_most_recent}
X.~Peng, S.~Huang, H.~Jiang, A.~Lu, and S.~Yu, ``{DNN+NeuroSim} v2.0: An end-to-end benchmarking framework for compute-in-memory accelerators for on-chip training,'' \emph{IEEE Transactions on Computer-Aided Design of Integrated Circuits and Systems}, vol.~40, no.~11, pp. 2306--2319, 2021.

\bibitem{mnsim_1}
L.~Xia, B.~Li, T.~Tang, P.~Gu, X.~Yin, W.~Huangfu, P.-Y. Chen, S.~Yu, Y.~Cao, Y.~Wang, Y.~Xie, and H.~Yang, ``{MNSIM}: Simulation platform for memristor-based neuromorphic computing system,'' in \emph{2016 Design, Automation \& Test in Europe Conference \& Exhibition (DATE)}, 2016, pp. 469--474.

\bibitem{mnsim_most_recent}
Z.~Zhu, H.~Sun, T.~Xie, Y.~Zhu, G.~Dai, L.~Xia, D.~Niu, X.~Chen, X.~S. Hu, Y.~Cao, Y.~Xie, H.~Yang, and Y.~Wang, ``{MNSIM} 2.0: A behavior-level modeling tool for processing-in-memory architectures,'' \emph{IEEE Transactions on Computer-Aided Design of Integrated Circuits and Systems}, vol.~42, no.~11, pp. 4112--4125, 2023.

\bibitem{accelergy}
Y.~N. Wu, J.~S. Emer, and V.~Sze, ``Accelergy: An architecture-level energy estimation methodology for accelerator designs,'' in \emph{2019 IEEE/ACM International Conference on Computer-Aided Design (ICCAD)}, 2019, pp. 1--8.

\bibitem{accelergy_pim}
Y.~N. Wu, V.~Sze, and J.~S. Emer, ``An architecture-level energy and area estimator for processing-in-memory accelerator designs,'' in \emph{2020 IEEE International Symposium on Performance Analysis of Systems and Software (ISPASS)}, 2020, pp. 116--118.

\bibitem{sparseloop_1}
Y.~N. Wu, P.-A. Tsai, A.~Parashar, V.~Sze, and J.~S. Emer, ``Sparseloop: An analytical, energy-focused design space exploration methodology for sparse tensor accelerators,'' in \emph{2021 IEEE International Symposium on Performance Analysis of Systems and Software (ISPASS)}, 2021, pp. 232--234.

\bibitem{sparseloop_2}
------, ``Sparseloop: An analytical approach to sparse tensor accelerator modeling,'' in \emph{2022 55th IEEE/ACM International Symposium on Microarchitecture (MICRO)}, 2022, pp. 1377--1395.

\bibitem{timeloop}
A.~Parashar, P.~Raina, Y.~S. Shao, Y.-H. Chen, V.~A. Ying, A.~Mukkara, R.~Venkatesan, B.~Khailany, S.~W. Keckler, and J.~Emer, ``Timeloop: A systematic approach to {DNN} accelerator evaluation,'' in \emph{2019 IEEE International Symposium on Performance Analysis of Systems and Software (ISPASS)}, 2019, pp. 304--315.

\bibitem{ruby}
\BIBentryALTinterwordspacing
M.~Horeni, P.~Taheri, P.~Tsai, A.~Parashar, J.~Emer, and S.~Joshi, ``Ruby: Improving hardware efficiency for tensor algebra accelerators through imperfect factorization,'' in \emph{2022 IEEE International Symposium on Performance Analysis of Systems and Software (ISPASS)}.\hskip 1em plus 0.5em minus 0.4em\relax Los Alamitos, CA, USA: IEEE Computer Society, may 2022, pp. 254--266. [Online]. Available: \url{https://doi.ieeecomputersociety.org/10.1109/ISPASS55109.2022.00039}
\BIBentrySTDinterwordspacing

\bibitem{NeuroSim_Validated}
A.~Lu, X.~Peng, W.~Li, H.~Jiang, and S.~Yu, ``{NeuroSim} validation with 40nm {RRAM} compute-in-memory macro,'' in \emph{2021 IEEE 3rd International Conference on Artificial Intelligence Circuits and Systems (AICAS)}, 2021, pp. 1--4.

\bibitem{jia}
H.~Jia, H.~Valavi, Y.~Tang, J.~Zhang, and N.~Verma, ``A programmable heterogeneous microprocessor based on bit-scalable in-memory computing,'' \emph{IEEE Journal of Solid-State Circuits}, vol.~55, no.~9, pp. 2609--2621, 2020.

\bibitem{sinangil}
M.~E. Sinangil, B.~Erbagci, R.~Naous, K.~Akarvardar, D.~Sun, W.-S. Khwa, H.-J. Liao, Y.~Wang, and J.~Chang, ``A 7-nm compute-in-memory sram macro supporting multi-bit input, weight and output and achieving 351 {TOPS/W} and 372.4 {GOPS},'' \emph{IEEE Journal of Solid-State Circuits}, vol.~56, no.~1, pp. 188--198, 2021.

\bibitem{wan}
W.~Wan, R.~Kubendran, S.~B. Eryilmaz, W.~Zhang, Y.~Liao, D.~Wu, S.~Deiss, B.~Gao, P.~Raina, S.~Joshi, H.~Wu, G.~Cauwenberghs, and H.-S.~P. Wong, ``33.1 a 74 {TMACS/W} {CMOS-RRAM} neurosynaptic core with dynamically reconfigurable dataflow and in-situ transposable weights for probabilistic graphical models,'' in \emph{2020 IEEE International Solid-State Circuits Conference - (ISSCC)}, 2020, pp. 498--500.

\bibitem{wan_ii}
\BIBentryALTinterwordspacing
W.~Wan, R.~Kubendran, C.~Schaefer, S.~B. Eryilmaz, W.~Zhang, D.~Wu, S.~Deiss, P.~Raina, H.~Qian, B.~Gao, S.~Joshi, H.~Wu, H.-S.~P. Wong, and G.~Cauwenberghs, ``\BIBforeignlanguage{en}{A compute-in-memory chip based on resistive random-access memory},'' \emph{\BIBforeignlanguage{en}{Nature}}, vol. 608, no. 7923, pp. 504--512, Aug. 2022, number: 7923 Publisher: Nature Publishing Group. [Online]. Available: \url{https://www.nature.com/articles/s41586-022-04992-8}
\BIBentrySTDinterwordspacing

\bibitem{wang}
H.~Wang, R.~Liu, R.~Dorrance, D.~Dasalukunte, D.~Lake, and B.~Carlton, ``A charge domain {SRAM} compute-in-memory macro with {C-2C} ladder-based 8-bit {MAC} unit in 22-nm {FinFET} process for edge inference,'' \emph{IEEE Journal of Solid-State Circuits}, vol.~58, no.~4, pp. 1037--1050, 2023.

\bibitem{wang_ii}
H.~Wang, R.~Liu, R.~Dorrance, D.~Dasalukunte, X.~Liu, D.~Lake, B.~Carlton, and M.~Wu, ``A 32.2 {TOPS/W} {SRAM} compute-in-memory macro employing a linear 8-bit {C-2C} ladder for charge domain computation in 22nm for edge inference,'' in \emph{2022 IEEE Symposium on VLSI Technology and Circuits (VLSI Technology and Circuits)}, 2022, pp. 36--37.

\bibitem{colonnade}
H.~Kim, T.~Yoo, T.~T.-H. Kim, and B.~Kim, ``Colonnade: A reconfigurable {SRAM}-based digital bit-serial compute-in-memory macro for processing neural networks,'' \emph{IEEE Journal of Solid-State Circuits}, vol.~56, no.~7, pp. 2221--2233, 2021.

\bibitem{dong_sram_cim}
Q.~Dong, M.~E. Sinangil, B.~Erbagci, D.~Sun, W.-S. Khwa, H.-J. Liao, Y.~Wang, and J.~Chang, ``15.3 a 351 {TOPS/W} and 372.4 {GOPS} compute-in-memory {SRAM} macro in 7nm {FinFET} {CMOS} for machine-learning applications,'' in \emph{2020 IEEE International Solid-State Circuits Conference-(ISSCC)}.\hskip 1em plus 0.5em minus 0.4em\relax IEEE, 2020, pp. 242--244.

\bibitem{xue202015}
C.-X. Xue, T.-Y. Huang, J.-S. Liu, T.-W. Chang, H.-Y. Kao, J.-H. Wang, T.-W. Liu, S.-Y. Wei, S.-P. Huang, W.-C. Wei \emph{et~al.}, ``15.4 a 22nm {2Mb} {ReRAM} compute-in-memory macro with 121-28{TOPS/W} for multibit mac computing for tiny {AI} edge devices,'' in \emph{2020 IEEE International Solid-State Circuits Conference-(ISSCC)}.\hskip 1em plus 0.5em minus 0.4em\relax IEEE, 2020, pp. 244--246.

\bibitem{dnn_scaling}
X.~Xu, Y.~Ding, S.~X. Hu, M.~Niemier, J.~Cong, Y.~Hu, and Y.~Shi, ``Scaling for edge inference of deep neural networks,'' \emph{Nature Electronics}, vol.~1, no.~4, pp. 216--222, 2018.

\bibitem{efficient_processing_of_dnns}
\BIBentryALTinterwordspacing
V.~Sze, Y.-H. Chen, T.-J. Yang, and J.~S. Emer, \emph{Efficient Processing of Deep Neural Networks}.\hskip 1em plus 0.5em minus 0.4em\relax Springer International Publishing, 2020. [Online]. Available: \url{https://doi.org/10.1007/978-3-031-01766-7}
\BIBentrySTDinterwordspacing

\bibitem{horowitz}
M.~Horowitz, ``1.1 computing's energy problem (and what we can do about it),'' in \emph{2014 IEEE International Solid-State Circuits Conference Digest of Technical Papers (ISSCC)}, 2014, pp. 10--14.

\bibitem{eyeriss}
Y.-H. Chen, T.~Krishna, J.~S. Emer, and V.~Sze, ``Eyeriss: An energy-efficient reconfigurable accelerator for deep convolutional neural networks,'' \emph{IEEE Journal of Solid-State Circuits}, vol.~52, no.~1, pp. 127--138, 2017.

\bibitem{PRIME}
P.~Chi, S.~Li, C.~Xu, T.~Zhang, J.~Zhao, Y.~Liu, Y.~Wang, and Y.~Xie, ``{PRIME}: A novel processing-in-memory architecture for neural network computation in reram-based main memory,'' in \emph{2016 ACM/IEEE 43rd Annual International Symposium on Computer Architecture (ISCA)}, 2016, pp. 27--39.

\bibitem{dot_product_engine}
M.~Hu, J.~P. Strachan, Z.~Li, E.~M. Grafals, N.~Davila, C.~Graves, S.~Lam, N.~Ge, J.~J. Yang, and R.~S. Williams, ``Dot-product engine for neuromorphic computing: Programming {1T1M} crossbar to accelerate matrix-vector multiplication,'' in \emph{2016 53nd ACM/EDAC/IEEE Design Automation Conference (DAC)}, 2016, pp. 1--6.

\bibitem{compute_in_dram}
M.~He, C.~Song, I.~Kim, C.~Jeong, S.~Kim, I.~Park, M.~Thottethodi, and T.~N. Vijaykumar, ``Newton: A {DRAM}-maker’s accelerator-in-memory {(AiM)} architecture for machine learning,'' in \emph{2020 53rd Annual IEEE/ACM International Symposium on Microarchitecture (MICRO)}, 2020, pp. 372--385.

\bibitem{compute_in_dram_ii}
S.~Xie, C.~Ni, A.~Sayal, P.~Jain, F.~Hamzaoglu, and J.~P. Kulkarni, ``16.2 {eDRAM}-{CIM}: Compute-in-memory design with reconfigurable embedded-dynamic-memory array realizing adaptive data converters and charge-domain computing,'' in \emph{2021 IEEE International Solid-State Circuits Conference (ISSCC)}, vol.~64, 2021, pp. 248--250.

\bibitem{ibm_14nm_pcm}
P.~Narayanan, S.~Ambrogio, A.~Okazaki, K.~Hosokawa, H.~Tsai, A.~Nomura, T.~Yasuda, C.~Mackin, S.~C. Lewis, A.~Friz, M.~Ishii, Y.~Kohda, H.~Mori, K.~Spoon, R.~Khaddam-Aljameh, N.~Saulnier, M.~Bergendahl, J.~Demarest, K.~W. Brew, V.~Chan, S.~Choi, I.~Ok, I.~Ahsan, F.~L. Lie, W.~Haensch, V.~Narayanan, and G.~W. Burr, ``Fully on-chip {MAC} at 14 nm enabled by accurate row-wise programming of {PCM}-based weights and parallel vector-transport in duration-format,'' \emph{IEEE Transactions on Electron Devices}, vol.~68, no.~12, pp. 6629--6636, 2021.

\bibitem{sttram_cim}
S.~Jain, A.~Ranjan, K.~Roy, and A.~Raghunathan, ``Computing in memory with spin-transfer torque magnetic {RAM},'' \emph{IEEE Transactions on Very Large Scale Integration (VLSI) Systems}, vol.~26, no.~3, pp. 470--483, 2018.

\bibitem{chen2022bit}
R.~Chen, H.~Kung, A.~Chandrakasan, and H.-S. Lee, ``A bit-level sparsity-aware {SAR} {ADC} with direct hybrid encoding for signed expressions for {AIoT} applications,'' in \emph{Proceedings of the ACM/IEEE International Symposium on Low Power Electronics and Design}, 2022, pp. 1--6.

\bibitem{CASCADE}
\BIBentryALTinterwordspacing
T.~Chou, W.~Tang, J.~Botimer, and Z.~Zhang, ``{CASCADE}: Connecting {RRAMs} to extend analog dataflow in an end-to-end in-memory processing paradigm,'' in \emph{Proceedings of the 52nd Annual IEEE/ACM International Symposium on Microarchitecture}, ser. MICRO '52.\hskip 1em plus 0.5em minus 0.4em\relax New York, NY, USA: Association for Computing Machinery, 2019, p. 114–125. [Online]. Available: \url{https://doi.org/10.1145/3352460.3358328}
\BIBentrySTDinterwordspacing

\bibitem{AtomLayer}
X.~Qiao, X.~Cao, H.~Yang, L.~Song, and H.~Li, ``{AtomLayer}: A universal {ReRAM}-based {CNN} accelerator with atomic layer computation,'' in \emph{2018 55th ACM/ESDA/IEEE Design Automation Conference (DAC)}, 2018, pp. 1--6.

\bibitem{raella}
\BIBentryALTinterwordspacing
T.~Andrulis, J.~S. Emer, and V.~Sze, ``{RAELLA}: Reforming the arithmetic for efficient, low-resolution, and low-loss analog {PIM}: No retraining required!'' in \emph{Proceedings of the 50th Annual International Symposium on Computer Architecture}, ser. ISCA '23.\hskip 1em plus 0.5em minus 0.4em\relax New York, NY, USA: Association for Computing Machinery, 2023. [Online]. Available: \url{https://doi.org/10.1145/3579371.3589062}
\BIBentrySTDinterwordspacing

\bibitem{1T2R_Aeris}
\BIBentryALTinterwordspacing
J.~Yue, Y.~Liu, F.~Su, S.~Li, Z.~Yuan, Z.~Wang, W.~Sun, X.~Li, and H.~Yang, ``{AERIS}: Area/energy-efficient {1T2R} reram based processing-in-memory neural network system-on-a-chip,'' in \emph{Proceedings of the 24th Asia and South Pacific Design Automation Conference}, ser. ASPDAC '19.\hskip 1em plus 0.5em minus 0.4em\relax New York, NY, USA: Association for Computing Machinery, 2019, p. 146–151. [Online]. Available: \url{https://doi.org/10.1145/3287624.3287635}
\BIBentrySTDinterwordspacing

\bibitem{TIMELY}
\BIBentryALTinterwordspacing
W.~Li, P.~Xu, Y.~Zhao, H.~Li, Y.~Xie, and Y.~Lin, ``{TIMELY}: Pushing data movements and interfaces in {PIM} accelerators towards local and in time domain,'' in \emph{Proceedings of the ACM/IEEE 47th Annual International Symposium on Computer Architecture}, ser. ISCA '20.\hskip 1em plus 0.5em minus 0.4em\relax IEEE Press, 2020, p. 832–845. [Online]. Available: \url{https://doi.org/10.1109/ISCA45697.2020.00073}
\BIBentrySTDinterwordspacing

\bibitem{teaal}
N.~Nayak, T.~O. Odemuyiwa, S.~Ugare, C.~W. Fletcher, M.~Pellauer, and J.~S. Emer, ``{TeAAL}: A declarative framework for modeling sparse tensor accelerators,'' 2023.

\bibitem{ResNet}
K.~He, X.~Zhang, S.~Ren, and J.~Sun, ``Deep residual learning for image recognition,'' \emph{2016 IEEE Conference on Computer Vision and Pattern Recognition (CVPR)}, pp. 770--778, 2016.

\bibitem{mars}
S.-H. Sie, J.-L. Lee, Y.-R. Chen, Z.-W. Yeh, Z.~Li, C.-C. Lu, C.-C. Hsieh, M.-F. Chang, and K.-T. Tang, ``{MARS}: Multimacro architecture {SRAM} {CIM}-based accelerator with co-designed compressed neural networks,'' \emph{IEEE Transactions on Computer-Aided Design of Integrated Circuits and Systems}, vol.~41, no.~5, pp. 1550--1562, 2022.

\bibitem{forms}
G.~Yuan, P.~Behnam, Z.~Li, A.~Shafiee, S.~Lin, X.~Ma, H.~Liu, X.~Qian, M.~N. Bojnordi, Y.~Wang, and C.~Ding, ``{FORMS}: Fine-grained polarized {ReRAM-based} in-situ computation for mixed-signal {DNN} accelerator,'' in \emph{2021 ACM/IEEE 48th Annual International Symposium on Computer Architecture (ISCA)}, 2021, pp. 265--278.

\bibitem{PIM-Prune}
C.~Chu, Y.~Wang, Y.~Zhao, X.~Ma, S.~Ye, Y.~Hong, X.~Liang, Y.~Han, and L.~Jiang, ``{PIM}-prune: Fine-grain {DCNN} pruning for crossbar-based process-in-memory architecture,'' in \emph{2020 57th ACM/IEEE Design Automation Conference (DAC)}, 2020, pp. 1--6.

\bibitem{learning_sparsity_for_reram}
\BIBentryALTinterwordspacing
J.~Lin, Z.~Zhu, Y.~Wang, and Y.~Xie, ``Learning the sparsity for {ReRAM}: Mapping and pruning sparse neural network for {ReRAM} based accelerator,'' in \emph{Proceedings of the 24th Asia and South Pacific Design Automation Conference}, ser. ASPDAC '19.\hskip 1em plus 0.5em minus 0.4em\relax New York, NY, USA: Association for Computing Machinery, 2019, p. 639–644. [Online]. Available: \url{https://doi.org/10.1145/3287624.3287715}
\BIBentrySTDinterwordspacing

\bibitem{fidelity_encoding_exploration}
T.~P. Xiao, B.~Feinberg, C.~H. Bennett, V.~Prabhakar, P.~Saxena, V.~Agrawal, S.~Agarwal, and M.~J. Marinella, ``On the accuracy of analog neural network inference accelerators [feature],'' \emph{IEEE Circuits and Systems Magazine}, vol.~22, pp. 26--48, 2021.

\bibitem{low_value_adc}
H.~Yun, H.~Shin, M.~Kang, and L.-S. Kim, ``Optimizing {ADC} utilization through value-aware bypass in {ReRAM}-based {DNN} accelerator,'' in \emph{2021 58th ACM/IEEE Design Automation Conference (DAC)}, 2021, pp. 1087--1092.

\bibitem{gpt2}
A.~Radford, J.~Wu, R.~Child, D.~Luan, D.~Amodei, I.~Sutskever \emph{et~al.}, ``Language models are unsupervised multitask learners,'' \emph{OpenAI blog}, vol.~1, no.~8, p.~9, 2019.

\bibitem{CACTI}
N.~P. Jouppi, A.~B. Kahng, N.~Muralimanohar, and V.~Srinivas, ``{CACTI-IO}: {CACTI} with off-chip power-area-timing models,'' \emph{IEEE Transactions on Very Large Scale Integration (VLSI) Systems}, vol.~23, no.~7, pp. 1254--1267, 2015.

\bibitem{aladdin}
Y.~S. Shao, B.~Reagen, G.-Y. Wei, and D.~Brooks, ``The aladdin approach to accelerator design and modeling,'' \emph{IEEE Micro}, vol.~35, no.~3, pp. 58--70, 2015.

\bibitem{adc_plug_in}
T.~Andrulis, R.~Chen, H.-S. Lee, J.~S. Emer, and V.~Sze, ``Modeling analog-digital-converter energy and area for compute-in-memory accelerator design,'' 2024.

\bibitem{adc_survey}
B.~Murmann, ``{{ADC} Performance Survey 1997-2023},'' [Online]. Available: \url{https://github.com/bmurmann/ADC-survey}.

\bibitem{ADC_scaling}
M.~Saberi, R.~Lotfi, K.~Mafinezhad, and W.~A. Serdijn, ``Analysis of power consumption and linearity in capacitive digital-to-analog converters used in successive approximation {ADCs},'' \emph{IEEE Transactions on Circuits and Systems I: Regular Papers}, vol.~58, no.~8, pp. 1736--1748, 2011.

\bibitem{ADC_Scaling_Murmann}
M.~Verhelst and B.~Murmann, ``Area scaling analysis of {CMOS} {ADCs},'' \emph{Electronics Letters}, vol.~48, pp. 314--315, 2012.

\bibitem{schreier}
R.~Schreier, J.~Silva, J.~Steensgaard, and G.~Temes, ``Design-oriented estimation of thermal noise in switched-capacitor circuits,'' \emph{IEEE Transactions on Circuits and Systems I: Regular Papers}, vol.~52, no.~11, pp. 2358--2368, 2005.

\bibitem{BRAHMS}
T.~Song, X.~Chen, X.~Zhang, and Y.~Han, ``{BRAHMS}: Beyond conventional {RRAM}-based neural network accelerators using hybrid analog memory system,'' in \emph{2021 58th ACM/IEEE Design Automation Conference (DAC)}, 2021, pp. 1033--1038.

\bibitem{cmos_scaling}
A.~Stillmaker and B.~Baas, ``Scaling equations for the accurate prediction of {CMOS} device performance from 180 nm to 7 nm,'' \emph{Integration}, vol.~58, pp. 74--81, 2017.

\bibitem{NeuroSim}
P.-Y. Chen, X.~Peng, and S.~Yu, ``{NeuroSim}+: An integrated device-to-algorithm framework for benchmarking synaptic devices and array architectures,'' in \emph{2017 IEEE International Electron Devices Meeting (IEDM)}, 2017, pp. 6.1.1--6.1.4.

\bibitem{imagenet}
J.~Deng, W.~Dong, R.~Socher, L.-J. Li, K.~Li, and L.~Fei-Fei, ``{ImageNet}: A large-scale hierarchical image database,'' in \emph{2009 IEEE Conference on Computer Vision and Pattern Recognition}, 2009, pp. 248--255.

\bibitem{cifar10}
A.~Krizhevsky, G.~Hinton \emph{et~al.}, ``Learning multiple layers of features from tiny images,'' 2009.

\bibitem{wiki:Bread}
Wikipedia, ``{Bread} --- {W}ikipedia{,} the free encyclopedia,'' \url{http://en.wikipedia.org/w/index.php?title=Bread&oldid=1189236003}, 2023, [Online; accessed 16-December-2023].

\bibitem{vit}
A.~Dosovitskiy, L.~Beyer, A.~Kolesnikov, D.~Weissenborn, X.~Zhai, T.~Unterthiner, M.~Dehghani, M.~Minderer, G.~Heigold, S.~Gelly, J.~Uszkoreit, and N.~Houlsby, ``An image is worth 16x16 words: Transformers for image recognition at scale,'' 2021.

\bibitem{mobilenetv3}
A.~Howard, M.~Sandler, G.~Chu, L.-C. Chen, B.~Chen, M.~Tan, W.~Wang, Y.~Zhu, R.~Pang, V.~Vasudevan, Q.~V. Le, and H.~Adam, ``Searching for {MobileNetV3},'' 2019.

\bibitem{fusion_i}
M.~Gilbert, Y.~N. Wu, A.~Parashar, V.~Sze, and J.~S. Emer, ``{LoopTree}: Enabling exploration of fused-layer dataflow accelerators,'' in \emph{2023 IEEE International Symposium on Performance Analysis of Systems and Software (ISPASS)}, 2023, pp. 316--318.

\bibitem{fusion_ii}
\BIBentryALTinterwordspacing
W.~Niu, J.~Guan, Y.~Wang, G.~Agrawal, and B.~Ren, ``{DNNFusion}: Accelerating deep neural networks execution with advanced operator fusion,'' in \emph{Proceedings of the 42nd ACM SIGPLAN International Conference on Programming Language Design and Implementation}, ser. PLDI 2021.\hskip 1em plus 0.5em minus 0.4em\relax New York, NY, USA: Association for Computing Machinery, 2021, p. 883–898. [Online]. Available: \url{https://doi.org/10.1145/3453483.3454083}
\BIBentrySTDinterwordspacing

\bibitem{PipeLayer}
L.~Song, X.~Qian, H.~Li, and Y.~Chen, ``{PipeLayer}: A pipelined {ReRAM}-based accelerator for deep learning,'' in \emph{2017 IEEE International Symposium on High Performance Computer Architecture (HPCA)}, 2017, pp. 541--552.

\bibitem{aihwkit}
M.~J. Rasch, D.~Moreda, T.~Gokmen, M.~Le~Gallo, F.~Carta, C.~Goldberg, K.~El~Maghraoui, A.~Sebastian, and V.~Narayanan, ``A flexible and fast {PyTorch} toolkit for simulating training and inference on analog crossbar arrays,'' in \emph{2021 IEEE 3rd International Conference on Artificial Intelligence Circuits and Systems (AICAS)}, 2021, pp. 1--4.

\bibitem{CrossSim}
\BIBentryALTinterwordspacing
T.~P. Xiao, C.~H. Bennett, B.~Feinberg, M.~J. Marinella, and S.~Agarwal, ``{CrossSim}: accuracy simulation of analog in-memory computing.'' [Online]. Available: \url{https://github.com/sandialabs/cross-sim}
\BIBentrySTDinterwordspacing

\bibitem{memtorch}
C.~Lammie and M.~R. Azghadi, ``{MemTorch}: A simulation framework for deep memristive cross-bar architectures,'' in \emph{2020 IEEE International Symposium on Circuits and Systems (ISCAS)}, 2020, pp. 1--5.

\bibitem{eva-cim}
D.~Gao, D.~Reis, X.~S. Hu, and C.~Zhuo, ``{Eva-CiM}: A system-level performance and energy evaluation framework for computing-in-memory architectures,'' \emph{IEEE Transactions on Computer-Aided Design of Integrated Circuits and Systems}, vol.~39, no.~12, pp. 5011--5024, 2020.

\bibitem{simeuro}
H.~Zhang, N.-M. Ho, D.~Y. Polat, P.~Chen, M.~Wahib, T.~T. Nguyen, J.~Meng, R.~S.~M. Goh, S.~Matsuoka, T.~Luo, and W.-F. Wong, ``Simeuro: A hybrid {CPU-GPU} parallel simulator for neuromorphic computing chips,'' \emph{IEEE Transactions on Parallel and Distributed Systems}, vol.~34, no.~10, pp. 2767--2782, 2023.

\bibitem{superneuro}
\BIBentryALTinterwordspacing
P.~Date, C.~Gunaratne, S.~R.~Kulkarni, R.~Patton, M.~Coletti, and T.~Potok, ``{SuperNeuro}: A fast and scalable simulator for neuromorphic computing,'' in \emph{Proceedings of the 2023 International Conference on Neuromorphic Systems}, ser. ICONS '23.\hskip 1em plus 0.5em minus 0.4em\relax New York, NY, USA: Association for Computing Machinery, 2023. [Online]. Available: \url{https://doi.org/10.1145/3589737.3606000}
\BIBentrySTDinterwordspacing

\bibitem{puma}
\BIBentryALTinterwordspacing
A.~Ankit, I.~E. Hajj, S.~R. Chalamalasetti, G.~Ndu, M.~Foltin, R.~S. Williams, P.~Faraboschi, W.-m.~W. Hwu, J.~P. Strachan, K.~Roy, and D.~S. Milojicic, ``{PUMA}: A programmable ultra-efficient memristor-based accelerator for machine learning inference,'' in \emph{Proceedings of the Twenty-Fourth International Conference on Architectural Support for Programming Languages and Operating Systems}, ser. ASPLOS '19.\hskip 1em plus 0.5em minus 0.4em\relax New York, NY, USA: Association for Computing Machinery, 2019, p. 715–731. [Online]. Available: \url{https://doi.org/10.1145/3297858.3304049}
\BIBentrySTDinterwordspacing

\bibitem{sparse_dnn}
C.~Guo, B.~Y. Hsueh, J.~Leng, Y.~Qiu, Y.~Guan, Z.~Wang, X.~Jia, X.~Li, M.~Guo, and Y.~Zhu, ``Accelerating sparse {DNN} models without hardware-support via tile-wise sparsity,'' in \emph{Proceedings of the International Conference for High Performance Computing, Networking, Storage and Analysis}, ser. SC '20.\hskip 1em plus 0.5em minus 0.4em\relax IEEE Press, 2020.

\bibitem{sun2023analog}
J.~Sun, P.~Houshmand, and M.~Verhelst, ``Analog or digital in-memory computing? benchmarking through quantitative modeling,'' in \emph{2023 IEEE/ACM International Conference on Computer Aided Design (ICCAD)}, 2023, pp. 1--9.

\bibitem{mei2021zigzag}
L.~Mei, P.~Houshmand, V.~Jain, S.~Giraldo, and M.~Verhelst, ``Zigzag: Enlarging joint architecture-mapping design space exploration for dnn accelerators,'' \emph{IEEE Transactions on Computers}, vol.~70, no.~8, pp. 1160--1174, 2021.

\bibitem{photonics}
T.~Andrulis, G.~I. Chaudhry, V.~M. Suriyakumar, J.~S. Emer, and V.~Sze, ``Architecture-level modeling of photonic deep neural network accelerators,'' in \emph{2024 IEEE International Symposium on Performance Analysis of Systems and Software (ISPASS)}, 2024.

\end{thebibliography}


\end{document}